\documentclass[aps,prx,twocolumn,showpacs,psfig,superscriptaddress,longbibliography]{revtex4-2}

\usepackage{times}
\usepackage{graphicx}
\usepackage{float}
\usepackage{latexsym,amsmath,amssymb,bm,euscript}
\usepackage{color}
\usepackage{subfigure}
\usepackage{epstopdf}
\usepackage[colorlinks=true,linkcolor=blue,citecolor=blue]{hyperref}
\usepackage{soul}
\usepackage[normalem]{ulem}
\usepackage{mathrsfs}
\usepackage{amsmath}
\usepackage{lettrine}
\usepackage{bbding}
\usepackage{xspace}
\usepackage{textcomp}
\usepackage{textcase}
\usepackage{setspace}
\usepackage{chemformula}
\usepackage{natbib}
\usepackage{tikz}

\newcommand{\cno}{\ch{CoNb_2O_6}}
\newcommand{\ya}{\ch{Yb$_4$As$_3$}}

\newcommand{\ccc}{\ch{Cs_2CuCl_4}}
\newcommand{\scvo}{\ch{SrCo_2V_2O_8}}
\newcommand{\bcvo}{\ch{BaCo_2V_2O_8}}

\newcommand{\crg}{\ch{CeRh_6Ge_4}}

\newcommand{\rev}[1]{{\color[rgb]{.0,.0,.0}{#1}}}

\def\para{\ensuremath{/\kern -0.8em /}\xspace}

\def\beqn{\begin{eqnarray}}
\def\eeqn{\end{eqnarray}}
\def\beq{\begin{equation}}
\def\eeq{\end{equation}}

\newcommand{\Beq}{\begin{eqnarray*} }
\newcommand{\Eeq}{\end{eqnarray*} }

\begin{document}

\title{Quantum Supercritical Regime with Universal Magnetocaloric Scaling in Ising Magnets}

\author{Enze Lv}
\affiliation{Institute of Theoretical Physics, Chinese Academy of Sciences, 
Beijing 100190, China}
\affiliation{School of Physical Sciences, University of Chinese Academy of 
Sciences, Beijing 100049, China}

\author{Ning~Xi}
\affiliation{Institute of Theoretical Physics, Chinese Academy of Sciences, 
Beijing 100190, China}

\author{Yuliang Jin}
\affiliation{Institute of Theoretical Physics, Chinese Academy of Sciences, 
Beijing 100190, China}

\author{Wei Li}
\email{w.li@itp.ac.cn}
\affiliation{Institute of Theoretical Physics, Chinese Academy of Sciences, 
Beijing 100190, China}

\begin{abstract} 
\end{abstract}
\date{\today}
\maketitle

\noindent{\bf{Abstract}}\\
Quantum critical points ubiquitously emerge in strongly correlated systems, with their influence persisting at finite temperatures and external fields. A paradigmatic example is the quantum Ising magnet, where transverse field $g$ controlling quantum fluctuations can expand the quantum critical point into an extended quantum critical regime. In this work, we propose a distinct quantum supercritical regime originating also from the quantum critical point but controlled by the longitudinal field $h$ coupled to the order parameter. Through thermal tensor network simulations, we find the quantum supercritical regime is enclosed by the finite-temperature crossover boundaries $T \propto h^{{z\nu}/\Delta}$, where $z$, $\nu$ and $\Delta \equiv \beta+\gamma$ are critical exponents. \rev{We comprehend} the supercritical scaling via thermal data collapse based on the derived scaling form.  Amongst other intriguing phenomena in quantum supercritical regime, there exists an enhanced magnetocaloric effect characterized by a universally diverging magnetic Gr\"uneisen ratio $\Gamma_h \propto T^{-\Delta/{z\nu}}$, \rev{which indicates that a} small symmetry-breaking field \rev{$h$} can generate dramatic temperature variation. We propose to observe the quantum supercritical regime in Ising-chain compound \cno~and \rev{related quantum materials, revealing a helium-3-free pathway to millikelvin cooling via the supercritical magnetocaloric effect.}
\\

\noindent{\bf{Introduction}}\\
Quantum critical point (QCP) embodies a scale-invariant many-body state that marks a continuous quantum phase transition between distinct phases of matter~\cite{Sachdev2000,Coleman2005,sachdev2015}. Nonclassical behaviors and universal scaling can be observed near the QCP, independent of microscopic details but determined by the universality class associated with global constraints such as symmetry and dimensionality. Although a QCP exists strictly at zero temperature, it profoundly shapes equilibrium and dynamical properties at finite temperatures~\cite{Sachdev2000,Coleman2005,sachdev2015}. The emergent phenomena linked to QCPs represent an intriguing topic in condensed matter physics~\cite{sondhi1997,Sachdev2000}, ultracold atoms~\cite{Xibo2012}, and quantum field theory~\cite{Sachdev2009,Mihailo2009}.

Specifically, a QCP extends into a quantum critical regime (QCR) at finite temperatures, where universal scaling behaviors emerge due to the intricate interplay between quantum and thermal fluctuations~\cite{Sachdev1992, sondhi1997, Sachdev2000, Coleman2005, Krichner2020, sachdev2015, continentino2017}. These finite-temperature scaling laws near QCP have been observed in various quantum materials. For instance, the Ising magnets \cno~\cite{coldea2010, kinross2014, wu2014, morris2014, liang2015, Fava2020, amelin2020, morris2021, xu2022, woodland2023, Ning2024}, \bcvo~\cite{Kimura2007, faure2018, wang2018, wang2019, zou2021, Wang2024Nature} and \scvo~\cite{He2006, Wang2015, Bera2017, cui2019, Wang2018Nature} exhibit quantum critical behaviors governed by the (1+1)D Ising universality class. These realistic systems serve as experimental platforms for exploring quantum criticality and its emergent phenomena. Among other notable phenomena, there exists a universal magnetocaloric effect (MCE) in the QCR. Through the adiabatic demagnetization process, unlike traditional paramagnetic cooling that reaches its lowest temperatures near zero field, the quantum critical cooling achieves its minimum temperature within the QCR. The fluctuation field $g$ --- such as the transverse field in the quantum Ising model --- controls quantum fluctuations and drives the QCP. A universal divergence of the magnetic Gr\"uneisen ratio, $\Gamma_g \equiv \frac{1}{T} (\frac{\partial T}{\partial g})_S \propto T^{-1/{z\nu}}$, with $z$ \rev{and} $\nu$ the critical exponents, has been theoretically predicted~\cite{zhu2003, Zhitomirsky2004, Markus2005, Honecker2009, Wu2011JPhCS, zhang2019} and experimentally confirmed in certain quantum magnets~\cite{Wolf2011, Wolf2014, Tokiwa2009, Tokiwa2015Signature, Gegenwart2016, Wolf2016, Oliver2017, Xiang2025}.

Notably, the role of symmetry-breaking fields --- like the longitudinal field $h$ in the quantum Ising magnets --- remains much less understood. In the ordered phase, field $h$ couples to the order parameter and induces the first-order phase transition line that terminates at a QCP, also known as quantum critical endpoint. The effective longitudinal field generated by interchain coupling --- for instance, in Ising-chain compounds like \cno~--- can give rise to massive quasi-particles described in terms of the exceptional \rev{$E_8$} Lie algebra~\cite{coldea2010, wu2014, morris2014, cui2019, amelin2020, zou2021, Ning2024}. However, the universal scaling behaviors induced by the longitudinal field near \rev{such QCP} remain to be systematically investigated. 

\begin{figure*}[t!]
\includegraphics[angle=0,width=1\linewidth]{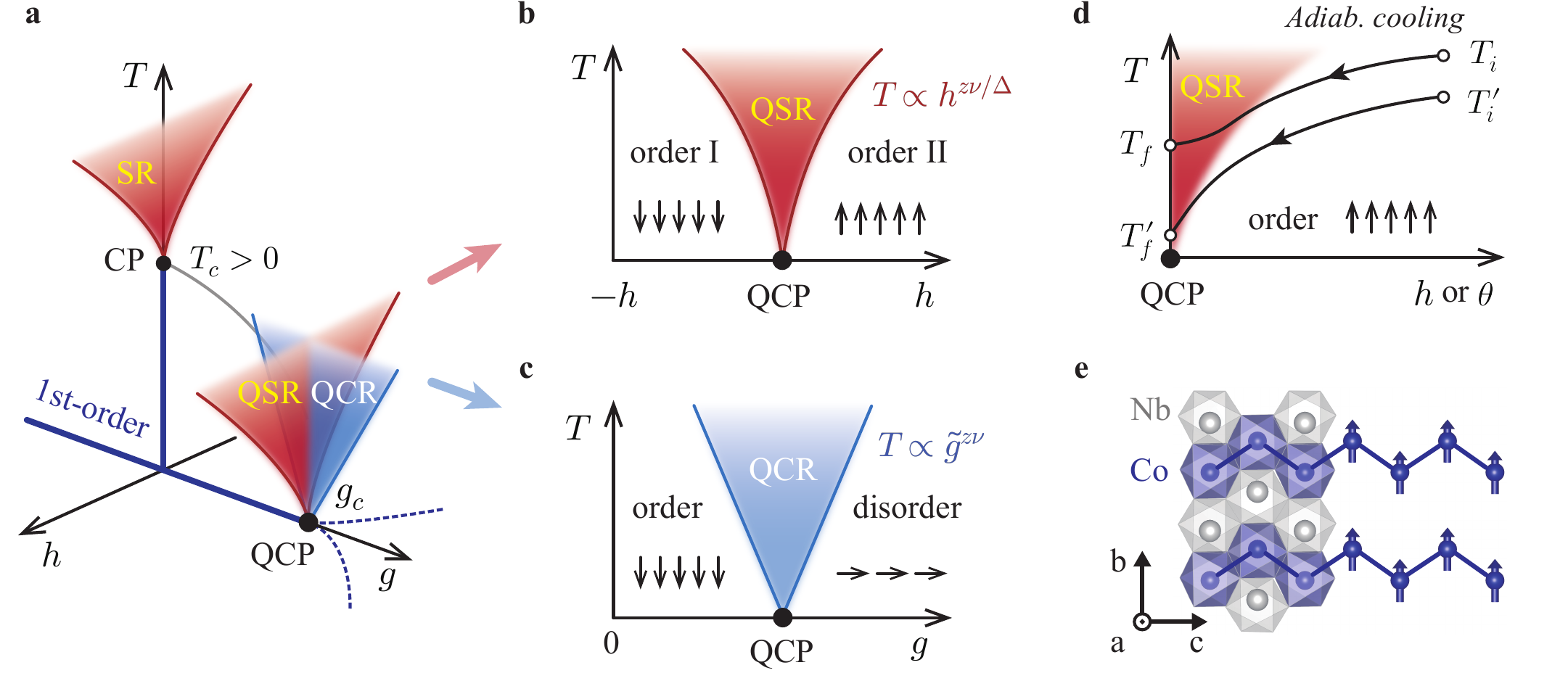}
\renewcommand{\figurename}{\textbf{Fig. }}
\caption{ \textbf{Schematic diagram of the quantum supercritical regime \rev{and universal magnetocaloric effects}.}
\textbf{a} Above the Curie point (CP) \(T_c\), there exists a classical supercritical regime (SR), reflecting the profound analogy between FM transitions and liquid-gas phase transitions. The SR is extended down by the quantum fluctuation $g$ and eventually becomes the quantum supercritical regime (QSR) originating from the quantum critical point (QCP). The QSR is located within the $h$-$T$ plane orthogonal to the quantum critical regime (QCR) situated in the $g$-$T$ plane. The two blue \rev{thick} lines represent the first-order phase transitions, and the \rev{blue curves} are quantum supercritical crossovers in the $g$-$h$ plane at zero temperature~\cite{wang2024qsc}. 
\textbf{b} The QSR is situated between two different ordered regimes in the $h$-$T$ plane, and the QSR crossover lines follow a distinct supercritical scaling $T \propto h^{{z\nu}/\Delta}$, with $z$, $\nu$ and $\Delta\equiv \beta+\gamma$ the critical exponents. 
\textbf{c} The QCR separates the ordered and disordered spin states in the $g$-$T$ plane. The two QCR crossover lines follow the quantum critical scaling $T \propto \tilde{g}^{z\nu}$,  where $\tilde{g} \equiv g-g_c$ measures the distance to the QCP at $g_c$. 
\textbf{d} Schematic diagram of quantum supercritical cooling. Starting from an ordered phase at initial temperature $T_i$ ($T_i'$), the quantum system reaches $T_f$ ($T_f'$) within the QSR by reducing or tilting the magnetic field ($\theta$ represents the field-tilting angle). 
\textbf{e} Crystal structure of \cno. The Co$^{2+}$ ions carry spin-1/2 moments and form Ising chains along the $c$ axis. 
}
\label{fig1}
\end{figure*}

We note that, in classical ferromagnetic (FM) systems, the symmetry-breaking field $h$ can extend a finite regime above the Curie temperature, as illustrated in Fig.~\ref{fig1}\textbf{a}. It is referred to as a classical supercritical regime (SR) first discovered in liquid-gas systems~\cite{Cagniard1822, Andrews1869}, where the supercritical fluids possess strong thermal fluctuations. Subsequent years witnessed intensive investigations of supercritical phenomena and crossover lines in the liquid-gas systems~\cite{xu2005, brazhkin2013, li2024supfluid}, which are very suggestive for the studies of magnetic phase transitions. \rev{We} now consider that the quantum-fluctuation field $g$ can suppress the Curie temperature $T_c$ in an FM Ising magnet and extend it down to absolute zero (c.f., Fig.~\ref{fig1}\textbf{a}). Given that, intriguing questions arise: Could there be a quantum supercritical regime (QSR) at low temperature, and whether it exhibits novel universal phenomena and distinct scaling behaviors as compared to QCR?

In this work, we unveil the existence of finite-temperature QSR --- emanating from the QCP and extended by the symmetry-breaking field $h$ --- through thermal tensor-network calculations on quantum spin systems~\cite{li2011, Dong2017bilayer, Chen2018XTRG, tanTRG2023}. Figure~\ref{fig1}\textbf{b} illustrates the QSR that resides in the $h$-$T$ plane, orthogonal to the QCR in the $g$-$T$ plane (Fig.~\ref{fig1}\textbf{c}). The QSR separates two distinct ordered phases, such as the spin-up and spin-down states in the Ising model, and the field $h$ couples to their order parameter. This leads to crossover lines characterized by a unique quantum supercritical scaling $T \propto h^{z\nu/\Delta}$. We further discover a pronounced MCE in the QSR, as shown in Fig.~\ref{fig1}\textbf{d}, characterized by universally diverging Gr\"uneisen ratio $\Gamma_h \equiv \frac{1}{T} (\frac{\partial T}{\partial h})_{S} \propto T^{-\Delta/{z\nu}}$ derived from the scaling analysis. Furthermore, we propose that quantum supercritical phenomena can be realized and investigated in \rev{Ising chain} \cno~\cite{coldea2010, kinross2014, wu2014, liang2015, Fava2020, amelin2020, morris2021, xu2022, woodland2023} (see Fig.~\ref{fig1}\textbf{e}) and \rev{other Ising} compounds~\cite{Kimura2007, faure2018, wang2018, wang2019, zou2021, Wang2024Nature, He2006, Wang2015, Bera2017, cui2019, Wang2018Nature, Xie2021Giant, Wendl2022Mesoscale, Liu2023Ultralow}, and design a field-tilting process for ultra-low temperature cooling. Additionally, we discuss other quantum supercritical \rev{systems}, including Heisenberg magnets with Dzyaloshinskii-Moriya (DM) interactions~\cite{Essler1998Dynamics, Oshikawa1997Gaps, Affleck1999Field, Zhao2003DMEffect, Lou2002Gap, Lou2005Midgap, Zvyagin2004, Oshikawa1999, Kohgi2001, Umegaki2015} and heavy-fermion materials~\cite{Bin2020}, etc, which serve as promising \rev{quantum material} coolants. \rev{Our findings establish a paradigm for exploring supercritical phenomena and demonstrate a route to extreme cooling via anisotropic quantum magnetic systems.} 
\\

\begin{figure*}[t!]
    \includegraphics[angle=0,width=1\linewidth]{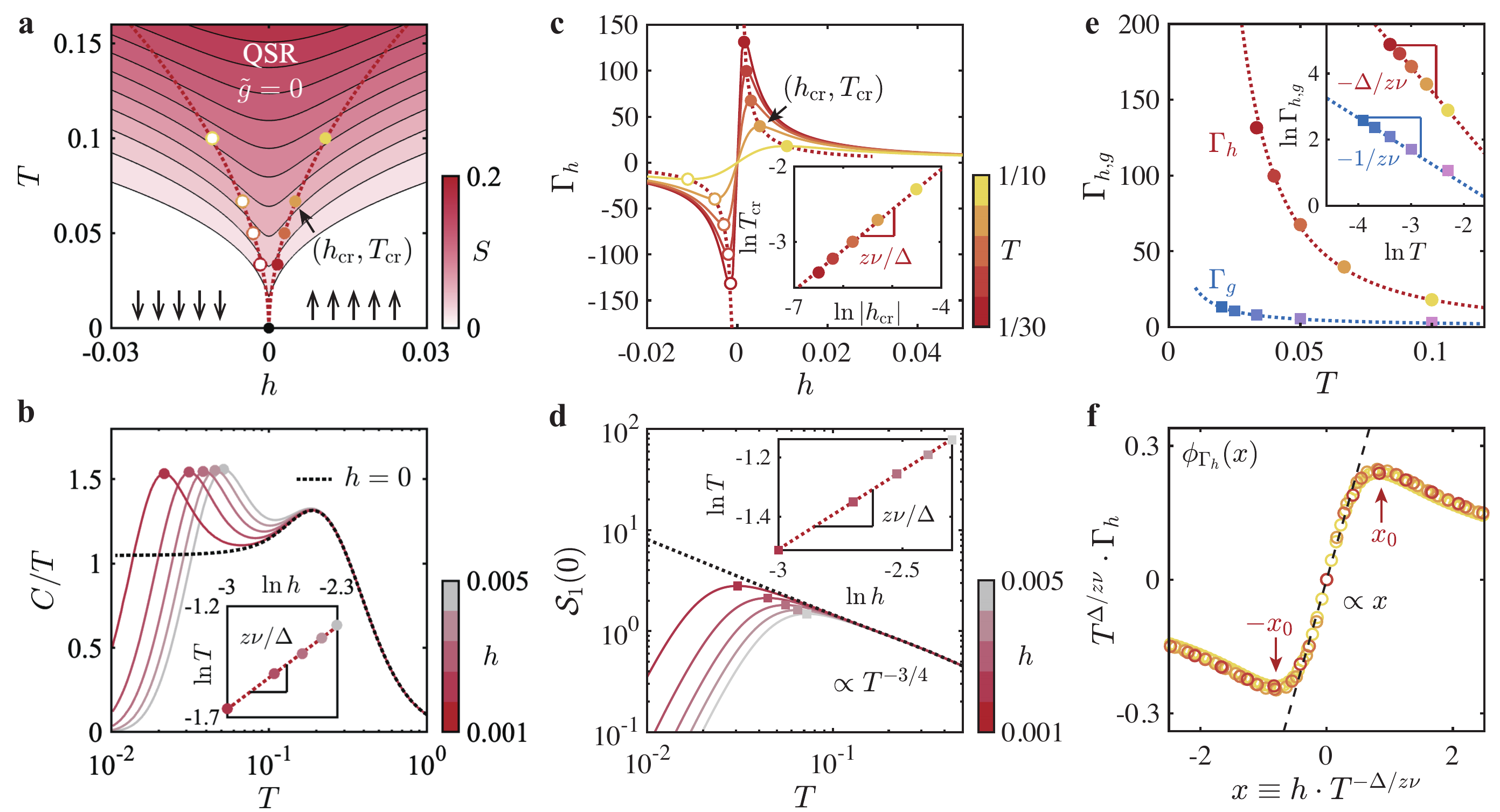}
    \renewcommand{\figurename}{\textbf{Fig. }}
    \caption{ \textbf{Universal quantum supercritical cooling with enhanced Gr\"uneisen ratio.}
    \textbf{a} The simulated isentropic lines of 1D quantum Ising model in the $h$-$T$ plane (with $\tilde{g}=0$). Red dashed lines represent the crossover lines enclosing the QSR. 
    \textbf{b} \rev{The double-peak specific heat curves for different field $h$. The position of the lower peak reveals the crossover line, scaling as $T_p \propto h^{{z\nu}/\Delta}$} (see inset). Solid circles in the main panel indicate the peaks of $C/T$, with one-to-one correspondence to those in the inset. 
    \textbf{c} The Gr\"uneisen ratio $\Gamma_h$ for each fixed temperature. Circles ($h_{\rm cr}, \rev{T}$), marking the peaks (dips) of $\Gamma_h$, constitute the crossover lines with an exponent ${z\nu}/\Delta$ (see inset). \rev{Red dashed lines mark the crossover boundary, as in panel \textbf{a}.}
    \textbf{d} The calculated spin-lattice relaxation rate $\mathcal{S}_1(\omega=0)$, which exhibits a power-law scaling $T^{-3/4}$ within the QSR, whose peak locations also reveal the QSR crossover line (see inset). Solid squares in the inset represent the peak locations \rev{$T_m$} in the main panel, \rev{which also exhbits the quantum supercritical scaling}. 
    \rev{Black dashed lines in \textbf{b} and \textbf{d} represent the $\tilde{g} = h = 0$ data.}
    \textbf{e} The comparison between peak values $\Gamma_h(\rev{h_{\rm cr}, T})$ of QSR and those of QCR (shown in Methods), where $\Gamma_{h}$ diverges much more rapidly as $\Delta =15/8 > 1$. \rev{The distinct scaling laws for both cases are illustrated by dashed lines.}
    \textbf{f} The scaling function $\phi_{\Gamma_h}(x)$ is obtained through data collapse from \textbf{c}, \rev{with color bar the same as \textbf{c}.} The $\pm x_0$ points correspond to the crossover lines shown in \textbf{a}, and the QSR is within the regime $-x_0 \leq x \leq x_0$. \rev{Black dashed line indicates the linear behavior near $x=0$.}
    }
    \label{fig2}
\end{figure*}

\noindent{\bf{Results}}\\
\textbf{Models.} 
We consider a representative spin-chain compound \cno, where the Ising interactions dominate and the quantum critical properties are effectively described by a 1D transverse-field Ising model~\cite{coldea2010, kinross2014, wu2014, Fava2020, amelin2020, woodland2023}. The Hamiltonian reads 
\begin{equation}
H/J = - \sum_{\langle i,j \rangle} S_i^z S_j^z - g \sum_i S_i^x - h \sum_i S_i^z,
\end{equation} 
where the Ising coupling $J \equiv 1$ is the energy scale, and $g$ ($h$) represents the transverse (longitudinal) field. The transverse field drives a QCP at $g_c=1/2$ between the FM and paramagnetic phases (see Fig.~\ref{fig1}\textbf{c}). Instead, the longitudinal field $h$ breaks the $\mathbb{Z}_2$ symmetry and induces a first-order quantum phase transition between two different FM states. 

To compute the low-temperature properties, we employ the thermal tensor network~\cite{li2011, Dong2017bilayer, Chen2018XTRG, tanTRG2023}, particularly the linearized and tangent-space tensor renormalization group methods  (see Methods and Supplementary Note 1). With these state-of-the-art approaches, we obtain the isentropic lines and magnetic Gr\"uneisen ratio characterizing the MCE with high accuracy.
\\

\noindent
\textbf{Quantum supercritical regime.} 
In Fig.~\ref{fig2}\textbf{a}, we present isentropic lines in the $h$-$T$ plane and highlight the QSR located between the two spin-ordered phases. The QSR and its crossover lines can be identified through thermodynamic properties and spin dynamics. In Fig.~\ref{fig2}\textbf{b}, the specific heat exhibits a double-peak structure, and the critical scaling $C \propto T^{d/z}$ (i.e., $\propto T$) appears at intermediate temperatures within the QSR. However, this scaling deviates at lower temperatures below the low-$T$ peaks of $C/T$, which determine the QSR crossover line in the $h$-$T$ plane (see inset of Fig.~\ref{fig2}\textbf{b}). These crossover lines follow the quantum supercritical scaling \rev{$T_{\rm p}\propto h^{z\nu/\Delta}$} (i.e., $\propto h^{8/15}$), governed by the critical exponents $z$, $\nu$ and $\Delta$ of the (1+1)D Ising universality class. 

Besides, \rev{the finite-temperature spin dynamics also exhibits quantum supercritical scaling.} We consider the spin-lattice relaxation rate in nuclear magnetic resonance, i.e., $1/T_1 \propto \mathcal{S}_1(\omega=0) \simeq \lim \limits_{\omega \rightarrow 0} T \sum_{\alpha=x,y,z} \chi_{\alpha \alpha}^{\prime \prime}(\omega)/\omega$, where $\chi_{\alpha \alpha}^{\prime \prime}(\omega)$ is the local dynamical susceptibility. With thermal tensor-network method, $1/T_1$ can be efficiently calculated via $\mathcal{S}_1(0) \simeq \frac{1}{T} \sum_\alpha \langle S_{i}^\alpha(\frac{1}{2T}) \, S_i^\alpha \rangle$~\cite{xi2024itp}. In Fig.~\ref{fig2}\textbf{d}, we uncover pronounced spin fluctuations within the QSR, characterized by a universally diverging $1/T_1 \propto T^{(\eta+d-2)/z} = T^{-3/4}$ with $\eta=1/4$, $d=1$, and $z=1$ for 1D quantum Ising chain. The quantity $\mathcal{S}_1(0)$ reaches a maximum at the QSR boundary and then becomes suppressed. The positions of these rounded peaks follow the scaling relation \rev{$T_{\rm m} \propto h^{z\nu/\Delta}$}, as demonstrated in Fig.~\ref{fig2}\textbf{d} and its inset.
\\

\begin{figure*}[t!]
\includegraphics[angle=0,width=0.7\linewidth]{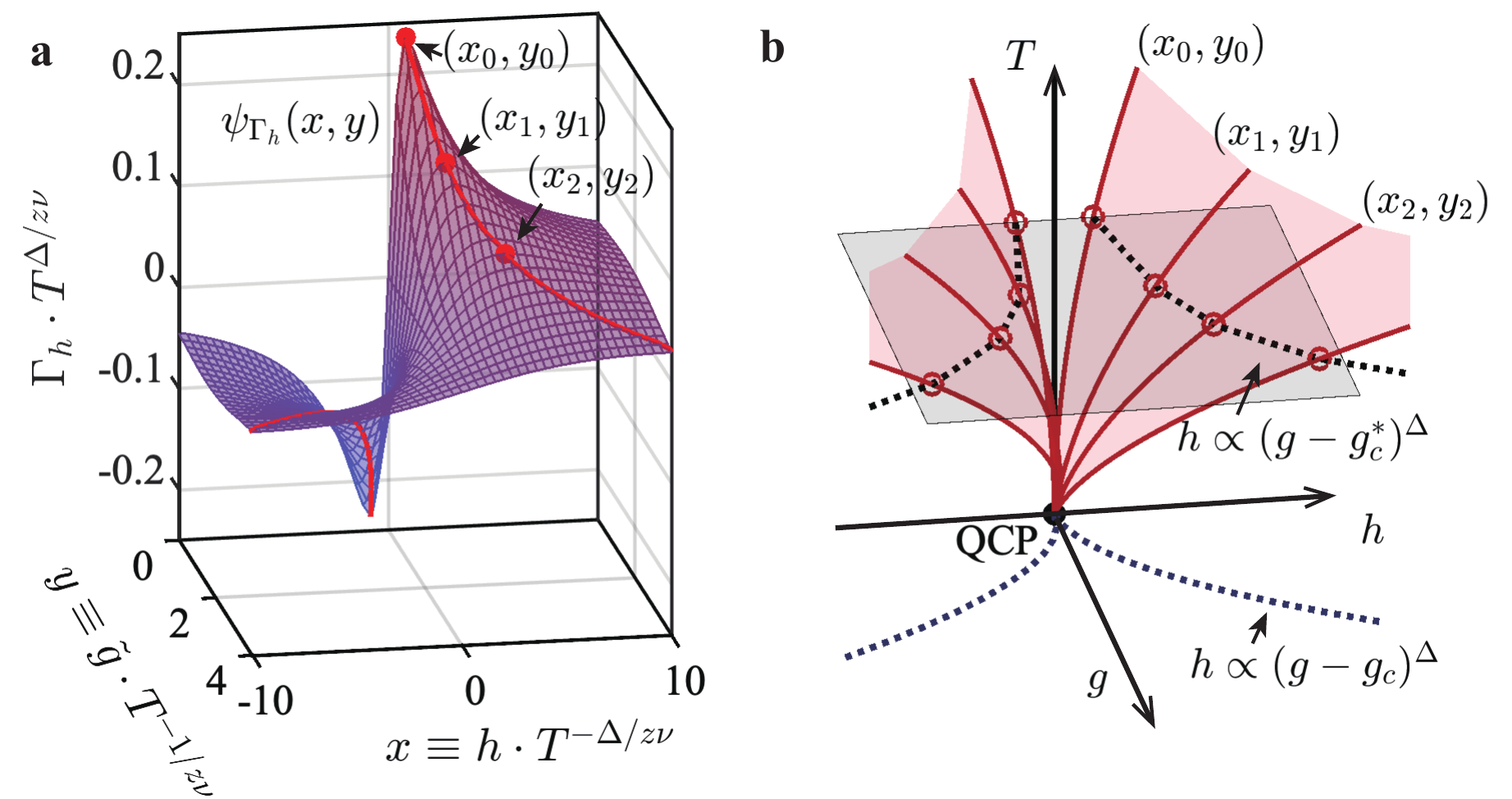}
\renewcommand{\figurename}{\textbf{Fig. }}
\caption{ \textbf{Hyperscaling function of Gr\"uneisen ratio and crossover surfaces near the QCP.}
\textbf{a} The hyperscaling function $\psi_{\Gamma_h} (x,y)$ near the QCP of quantum Ising model, obtained through data collapsing. The red lines indicate the locations of peaks/dips in $\psi_{\Gamma_h}(x,y)$ landscape when scanning $x$ for various fixed $y$. Each point $(x_i, y_i)$ in \textbf{a} corresponds to a crossover line in \textbf{b}, and \rev{these} lines form a crossover surface in the $g$-$h$-$T$ diagram. The gray plane represents an isothermal cut, where the dashed intersection lines connect the locations of \rev{the} maxima in $\Gamma_h$ at fixed $T$. In the low-temperature limit, they satisfy $h \propto (g - g_c)^{\Delta}$, i.e., quantum supercritical crossovers in the $g$-$h$ plane proposed in Ref.~\cite{wang2024qsc}.
}
\label{fig3}
\end{figure*}

\noindent
\textbf{Universal quantum supercritical cooling.} 
In Fig.~\ref{fig1}\textbf{d}, we have shown a schematic illustration of the pronounced quantum supercritical cooling, where low temperatures are achieved in the QSR. In Fig.~\ref{fig2}\textbf{a}, we show the simulated isentropic lines near the Ising QCP and reveal a significantly enhanced MCE, where even a small symmetry-breaking field $h$ can induce a dramatic temperature variation. To characterize the quantum supercritical cooling effect, we present the magnetic Gr\"uneisen ratio $\Gamma_h$ for each fixed temperature $T$ in Fig.~\ref{fig2}\textbf{c}. In the $h$-$T$ plane, $\Gamma_h$ exhibits a peak-dip structure with a sign change as the field $h$ crosses above the QCP. In the inset, the peak locations \rev{follow the quantum supercritical scaling law and represent} the QSR crossover lines in Fig.~\ref{fig2}\textbf{a}.

When comparing universal MCEs in the QSR and QCR~\cite{zhu2003}, there are notable differences and similarities. As shown in Fig.~\ref{fig2}\textbf{e}, the peak values $\Gamma_h(h_{\rm cr}, \rev{T})$ are orders of magnitude \rev{larger} than those of $\Gamma_g$ in the QCR at low temperatures. This can be explained with the distinct scaling laws demonstrated in the inset, where we can identify both the quantum supercritical scaling $\Gamma_h \propto T^{-\Delta/{z\nu}}$ {(i.e., $\propto T^{-15/8}$)} and critical scaling $\Gamma_g \propto T^{-1/{z\nu}}$ (i.e., $\propto T^{-1}$)~\cite{zhu2003, Markus2005, Zhitomirsky2004}, \rev{as confirmed by scaling analysis (see Methods)}. As $\Delta>1$ for Ising universality classes (and other conventional cases, see Tab.~\ref{Tab:Exponents}), the quantum supercritical scaling characterizes a more pronounced MCE. Beyond 1D chain, we also find very prominent QSR cooling in 2D Ising model (see Supplementary Note \rev{2}).

To understand the quantum supercritical scaling law, we perform data collapse on Gr\"uneisen ratio $\Gamma_h$ using the derived scaling form $\Gamma_h = T^{-\Delta/{z\nu}} \phi_{\Gamma_h}(x)$, where $\phi_{\Gamma_h}(x)$ is the scaling function with $x \equiv h \cdot T^{-\Delta/{z\nu}}$ (see detailed scaling analysis in Methods). As shown in Fig.~\ref{fig2}\textbf{f}, we obtain the scaling function $\phi_{\Gamma_h}(x)$ from an excellent collapse of the calculated Gr\"uneisen ratio data in Fig.~\ref{fig2}\textbf{c}. $\phi_{\Gamma_h}(x)$ is a parity-odd, smooth function, displaying a peak and a dip similar to Fig.~\ref{fig2}\textbf{c}, but without any singularity. Given the scaling form, we have $\partial \Gamma_h / \partial h = T^{-2\Delta/z\nu} \phi_{\Gamma_h}'(x) = 0$, indicating that the extreme values of isothermal $\Gamma_h$ correspond to the zero points $x=\pm x_{0}$ of $\phi_{\Gamma_h}'(x)$, as marked by the arrows in Fig.~\ref{fig2}\textbf{f}. This naturally leads to the QSR crossover lines $h_{\rm cr} = \pm x_{0} \cdot \rev{T}^{\Delta/{z\nu}}$ (see, e.g., Fig.~\ref{fig2}\textbf{a}). Moreover, as $\phi_{\Gamma_h}(\pm x_0)$ is a finite constant in the scaling function, we \rev{find} that $\Gamma_h = T^{-\Delta/{z\nu}}\phi_{\Gamma_h}(\pm x_{0}) \propto T^{-\Delta/{z\nu}}$, explaining the observed supercritical scaling law in Fig.~\ref{fig2}\textbf{e}. Lastly, in the QSR ($|x|\le x_0$), $\phi_{\Gamma_h}(x)$ is approximately a linear function, i.e., $\Gamma_h \cdot T^{\Delta/{z\nu}} \propto x$ (see Fig.~\ref{fig2}\textbf{f}). Recalling that $x \equiv h \cdot T^{-\Delta/{z\nu}}$, we have that $\Gamma_h/h \propto T^{-2\Delta/{z\nu}}$, demonstrating a even stronger divergence within the QSR and for fixed $h$ (see Methods and Supplementary Note \rev{3}). This scaling law further underscores the enhanced quantum supercritical cooling. 
\\

\noindent
\textbf{Hyperscaling function and crossover surfaces.} 
{More generally,  we consider the simultaneous effects of $g$ and $h$ fields near the QCP. In this case,} the Gr\"uneisen ratio $\Gamma_h$ can be described as a two-variable hyperscaling function, $\Gamma_h = T^{-\Delta/{z\nu}} \psi_{\Gamma_h}(x, y)$, with $x$ defined above and $y\equiv \tilde{g} \cdot T^{-1/{z\nu}}$. In Fig.~\ref{fig3}\textbf{a}, we show the obtained function $\psi_{\Gamma_h}(x,y)$ through data collapsing. The red solid lines mark the peaks and dips of $\psi_{\Gamma_h}(x,y)$ by scanning $x$ for fixed $y$, and each point $(x_i, y_i)$ represents a crossover line in Fig.~\ref{fig3}\textbf{b}. For example, the point with {$(x_0,y_0=0)$} corresponds to the QSR crossover line for $\tilde{g}=0$ in Fig.~\ref{fig2}\textbf{a}. As $(x_i, y_i)$ moves, these crossover lines form a crossover surface in Fig.~\ref{fig3}\textbf{b}, enclosing a generalized QSR bulk for $\tilde{g} \neq 0$. Notably, an isothermal plane at a fixed temperature $T$ (gray plane) intersects the crossover surfaces, resulting in two dashed lines that follow the scaling $h \propto (g - g_c^*)^{\Delta}$, with $g_c^*~\rev{\lesssim}~g_c$. In the zero-temperature limit, $g_c^* = g_c$, and the dashed line in Fig.~\ref{fig1}\textbf{a} is nothing but the ground-state supercritical crossover line $h \propto \tilde{g}^{\Delta}$ proposed in Ref.~\cite{wang2024qsc} (see Supplementary Note \rev{4}).
\\

\begin{figure*}[t!]
\includegraphics[angle=0,width=0.7\linewidth]{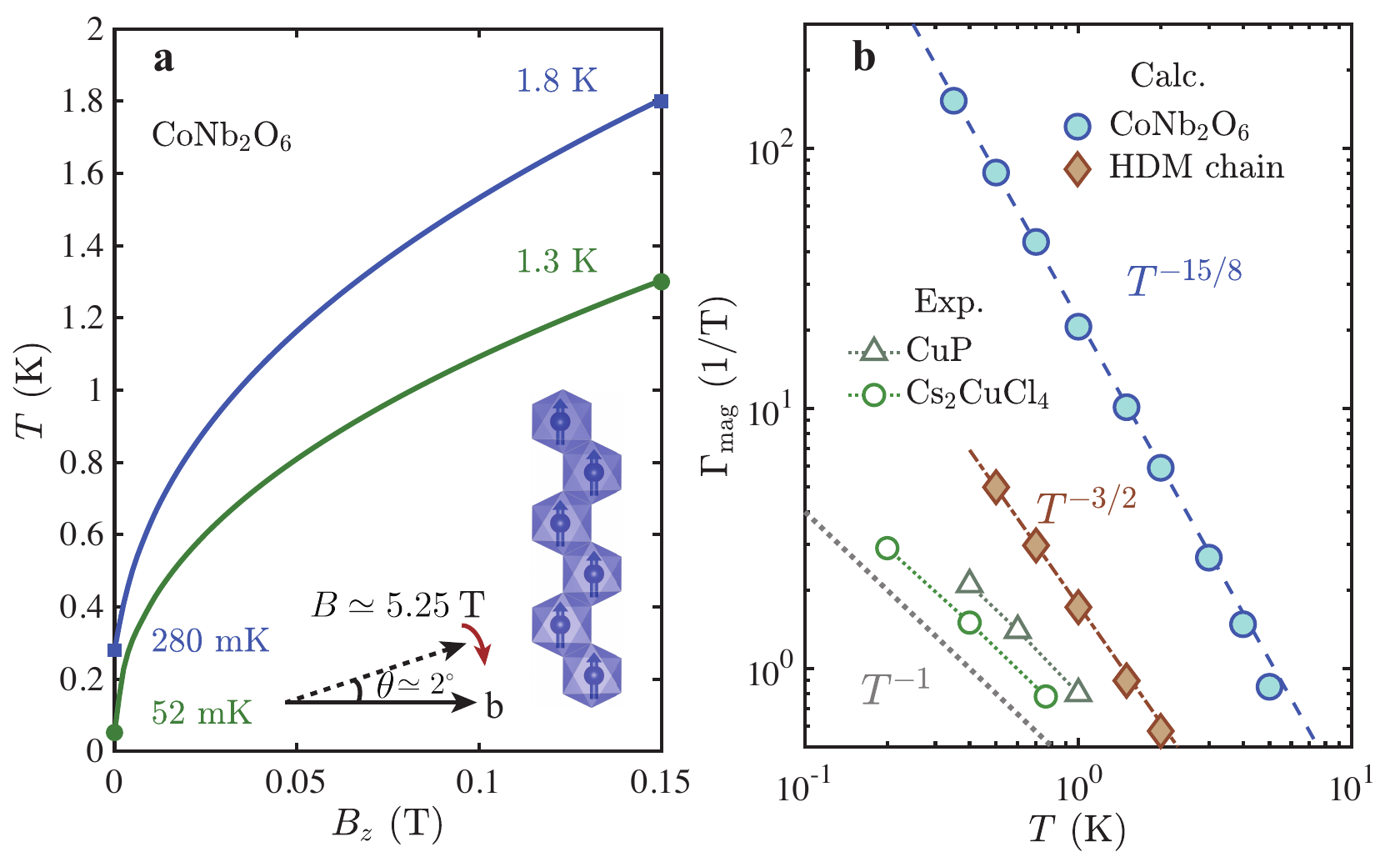}
\renewcommand{\figurename}{\textbf{Fig. }}
\caption{ \textbf{Quantum supercritical cooling effect for the Ising magnet \cno.}
\textbf{a} Simulated isentropic lines of \cno~by varying the longitudinal field $B_z$, while the transverse field is fixed at $B_x^c \simeq 5.25$~T. The simulations are performed using realistic parameters for the compound, with results presented in experimental units. Inset shows the field-tilting protocol for observing quantum supercritical behaviors in \cno. A small tilting angle of about 2$^\circ$ generates sufficient $B_z$ component to enable quantum supercritical cooling into the ultralow-temperature regime. 
\textbf{b} Temperature scaling of magnetic Gr\"uneisen peak values for various systems. Blue circles represent the calculated results of \cno. The supercritical Heisenberg-Dzyaloshinskii-Moriya (HDM) system is also plotted as a comparison, where the coupling strength is set as $J\simeq 18.2$~K as in the compound Cu benzoate. The hollow dots are adapted from experimental results of different materials, such as the triangular-lattice Heisenberg antiferromagnet \ccc~\cite{Wolf2014} and \rev{the} spin chain \ch{CuP}~\cite{Oliver2017}. 
}
\label{fig4}
\end{figure*}

\noindent
\textbf{Quantum supercriticality in realistic materials.}
The Ising magnet {\cno} has an FM interaction $J\simeq 28.8$~K, with a QCP at the critical transverse field of $B_x^c \simeq 5.25$~T~\cite{coldea2010, kinross2014, liang2015, amelin2020, morris2021, xu2022}. In Fig.~\ref{fig4}\textbf{a}, we calculate the universal scaling in {\cno} and demonstrate that even a small longitudinal field can induce significant temperature variations extending into the deeply sub-Kelvin regime. Remarkably, starting from an initial temperature 1.8~K, \cno~cools to 280~mK under a very small change of longitudinal field, while from 1.3~K, it achieves an ultralow temperature 52~mK, as shown in Fig.~\ref{fig4}\textbf{a}. The pronounced sensitivity to longitudinal magnetic fields facilitates the implementation of a field-tilting protocol to probe quantum supercritical phenomena, characterize their scaling properties, and observe the universal cooling effect.

To be specific, by slightly tilting the magnetic fields, as shown in the inset of Fig.~\ref{fig4}\textbf{a}, a small longitudinal field is introduced, while keeping the transverse component largely unchanged. A small field-tilting angle (approximately $2^\circ$) maintains the transverse field at $g \simeq 5.25$~T near the QCP, while the longitudinal component $h \simeq 0.15$~T induces a pronounced quantum supercritical cooling effect in \cno. This field-tilting cooling approach is experimentally viable, as demonstrated by recent studies of 3D Ising magnet LiREF$_4$~\cite{Xie2021Giant,Wendl2022Mesoscale,Liu2023Ultralow}. While large entropy changes are observed in those studies with ultralow field variation, the quantum supercriticality and universal magnetocaloric scaling predicted in the current study still need to be investigated in experiments.

To compare the quantum critical and supercritical cooling, we collect the peak values in the Gr\"uneisen ratios by scanning the field at various fixed temperatures, and show the results in Fig.~\ref{fig4}\textbf{b}. From the comparisons, we find the Ising compound \cno~demonstrates a quantum supercritical scaling $\Gamma_B \propto T^{-\Delta/{z\nu}} = T^{-15/8}$, significantly exceeding the quantum critical divergence $\Gamma_B \propto T^{-1/{z\nu}} = T^{-1}$ observed in other compounds~\cite{Wolf2014,Oliver2017}. Besides, for the Ising-chain compound \ch{CoNb_2O_6}, additional interactions like the interchain coupling can lead to a low-temperature 3D ordered phase, where signatures of \rev{the} $E_8$ or $D_8^{(1)}$ spectrum are observed~\cite{coldea2010, wu2014, morris2014, cui2019, amelin2020, zou2021, Ning2024}.  Although the 1D QCP is embedded within a dome of 3D magnetic order in \cno~\cite{liang2015}, our predicted universal behaviors remain experimentally observable within a finite-temperature window that extends beyond the 3D ordered phase boundary. Consequently, we conclude that the experimental predictions regarding quantum supercritical scaling and its associated enhanced cooling effects remain robust against inter-chain perturbations. Besides, we present results for another representative supercritical system --- the Heisenberg chain with Dzyaloshinskii-Moriya (DM) interaction --- in Fig.~\ref{fig4}\textbf{b}, where a $T^{-3/2}$ scaling is observed, also surpassing the $T^{-1}$ scaling (see Methods and Supplementary Note 5).
\\

\noindent
{\bf{Discussion}}\\
The discovery of MCE originates from the pioneering work of Weiss and Piccard in 1917~\cite{Weiss1917, Smith2013}, who first reported ``very sensitive temperature variations in relation to the magnetic field''  above the Curie point $T_c$ of nickel. From a contemporary viewpoint, the MCE emerges as a distinctive supercritical phenomenon, since thermal fluctuations are most prominent above $T_c$ and rapidly attenuate below it. To extend the MCE to lower temperatures constitutes a significant challenge for cryogenic technologies. Recent studies have proposed utilizing frustrated quantum magnets as potential platforms for extreme cooling~\cite{Xiang2024Nature, Li2024TopoCooling}. 

In this study, we demonstrate that the quantum suppression of the finite-temperature Curie point to a zero-temperature QCP (i.e., a quantum critical endpoint in the $g$-$h$ plane) removes the limitation of a finite $T_c$ \rev{in magnetic cooling}. It further results in a universal Gr\"uneisen ratio $\Gamma_h$ that diverges with a significantly stronger exponent. Specifically, the divergence follows $T^{-\Delta/z\nu}$ in the quantum case, compared to $(T-T_c)^{-\Delta+1}$ in the classical case (see Supplementary Note \rev{2} for a detailed comparison). \rev{The quantum supercritical MCE thus opens possibilities for millikelvin refrigeration.}

Although the QSR proposed here and the well-studied QCR both serve as powerful analytical lenses for QCP and enable universal cooling effect, they are fundamentally distinct on multiple levels. The QSR and QCR exist in different planes, i.e., $h$-$T$ vs. $g$-$T$ plane, driven by the symmetry-breaking field $h$ and fluctuation field $g$, respectively. Thus the QSR separates two different ordered regimes breaking the same symmetry, while the QCR is located between the ordered and quantum disordered regimes. Their crossover boundaries also follow distinct scaling laws, i.e., $T \propto h^{{z\nu}/\Delta}$ (QSR) vs. $T \propto \tilde{g}^{z\nu}$ (QCR). Remarkably, the magnetic Gr\"uneisen ratio in the QSR exhibits a boosted power-law divergence, $\Gamma_h \propto T^{-\Delta/z\nu}$, where the exponent $\Delta > 1$ for common universality classes (see Tab.~\ref{Tab:Exponents} in Methods). This represents a dramatic amplification over the QCR scaling $\Gamma_g \propto T^{-1/z\nu}$. This enables orders-of-magnitude stronger field-driven thermal responses, as demonstrated in Fig.~\ref{fig4} by our calculations using realistic material parameters. 

Quantum supercritical phenomena can emerge across a variety of realistic quantum materials. We have thoroughly discussed above the experimental feasibility and field-tilting setup using the Ising magnet \cno. Furthermore, both the 2D Ising magnet \ch{TmMgGaO_4}~\cite{Li2020,Hu2020Exp} and 3D Ising magnet~\ch{LiHoF_4}~\cite{Xie2021Giant, Wendl2022Mesoscale, Liu2023Ultralow} demonstrate anisotropic thermodynamic responses in experiments, suggesting promising candidates for investigating quantum supercritical behaviors. Quantum magnets with DM interactions also exhibit quantum supercriticality;  \rev{there exists a series of candidates} with different coupling strengths --- including Cu benzoate~\cite{Oshikawa1997Gaps, Essler1998Dynamics, Affleck1999Field, Zhao2003DMEffect, Lou2002Gap, Lou2005Midgap}, Cu-PM~\cite{Zvyagin2004}, \ya~\cite{Oshikawa1999,Kohgi2001} and KCuGaF$_6$~\cite{Umegaki2015}. 

Beyond Mott insulator magnets, quantum supercriticality may also emerge in \rev{metallic ferromagnets~\cite{Brando2016}. A prominent example is} \crg, which hosts a pressure-induced FM QCP at $p_c\approx 0.8$~GPa~\cite{Bin2020} \rev{--- possibly as the endpoint of a first-order line}. Consequently, a uniform magnetic field --- coupling to the FM order parameter --- \rev{is expected to} induce quantum supercritical phenomena and generate significant MCE in such a heavy-fermion ferromagnet.

Overall, our finding not only establishes a theoretical framework for quantum supercriticality \rev{in Ising-type quantum materials, but also proposes an approach for quantum magnetic} cooling that surpasses conventional limits. Besides magnetocalorics, \rev{the strong quantum fluctuations and extreme sensitivity of the QSR may give rise to novel phenomena in spin dynamics~\cite{wang2024qsc}, magnetostriction, and even transport properties~\cite{Gao2025SSE}, etc,} offering a promising frontier for future exploration.
\\

\noindent{\bf{Methods}}\\
\textbf{Thermal tensor network methods.}
To accurately simulate the low-temperature properties of quantum lattice models, we employ the linearized tensor renormalization group (LTRG)~\cite{li2011, Dong2017bilayer} and tangent-space tensor renormalization group (tanTRG) methods~\cite{tanTRG2023, Qu2024Cuprate}. For Ising chains discussed in the main text, LTRG is exploited to simulate the infinite-chain system, and the calculated results exhibit excellent convergence and high accuracy. For example, with a bond dimension of $D=200$ and Trotter step of $\tau=0.01$, the relative error in specific heat is very small ($ \lesssim 10^{-3}$) at temperatures down to $T/J = 0.01$. Furthermore, we employ the tanTRG method to study the 2D quantum Ising model \rev{near the field-driven QCP. The calculations are performed} on cylindrical geometries with widths up to 12 sites (see Supplementary Notes 1 and \rev{2}).\\

\noindent
\textbf{Critical scaling analysis of $\Gamma_g$.}
Scaling functions serve as powerful tools for investigating universal behaviors and scaling laws. Here, we employ scaling forms to investigate the quantum critical cooling induced by transverse fields near the QCP. Previous studies have established a general scaling law for the Gr\"uneisen ratio at the QCP, i.e., $\Gamma_g(g_c,T)\propto T^{-1/z\nu}$~\cite{zhu2003, Zhitomirsky2004, Markus2005, Honecker2009, Wu2011JPhCS, zhang2019}. However, for the transverse-field Ising chain, the Gr\"uneisen ratio remains a constant, i.e., $\Gamma_g(g_c,T) = 1/2$ at the self-dual QCP~\cite{zhang2019}. This non-diverging $\Gamma_g(g_c,T)$ at QCP is thus no longer appropriate for characterizing the quantum critical cooling in Ising chain. 

\begin{figure}[h!]
    \includegraphics[angle=0,width=1\linewidth]{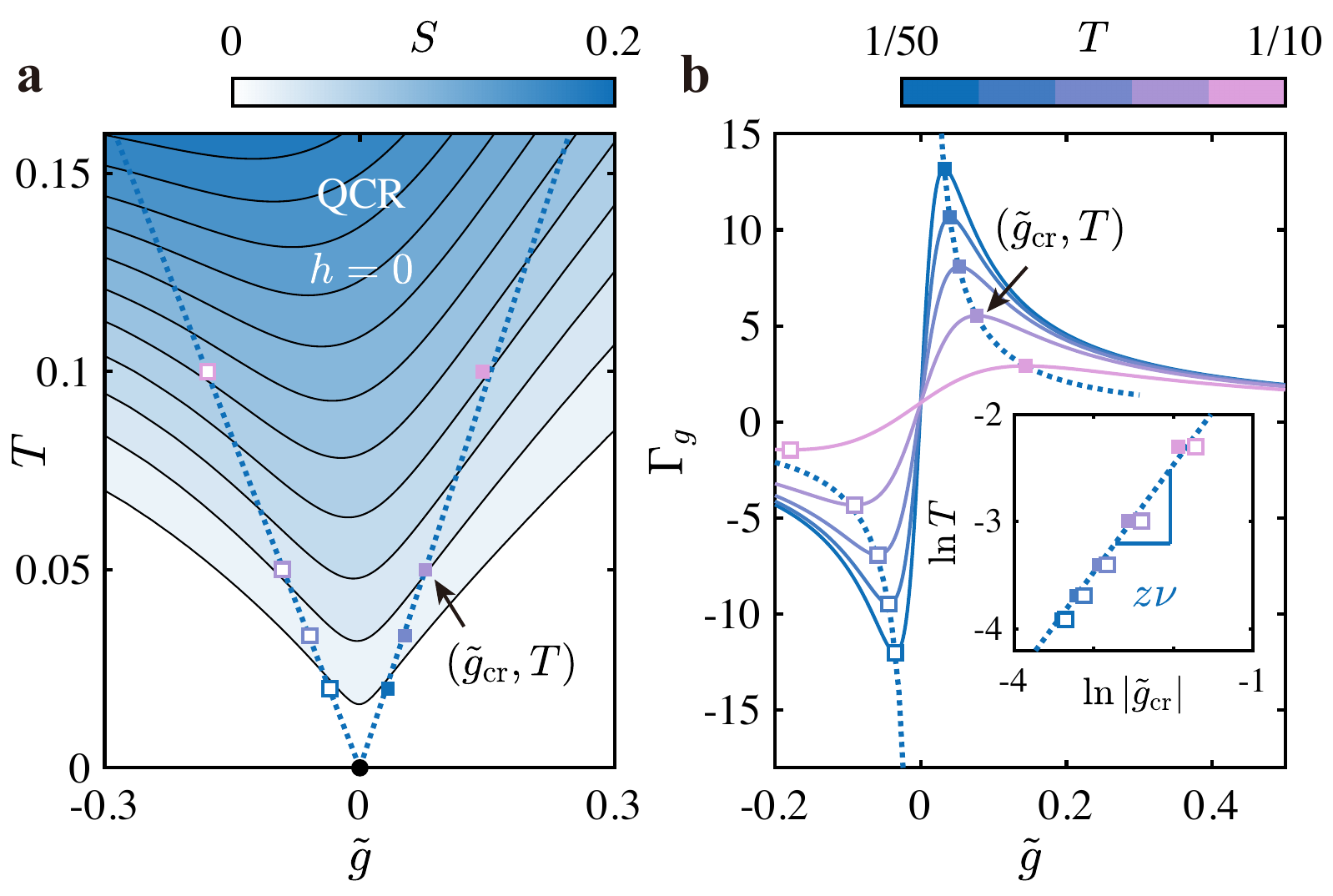}
    \renewcommand{\figurename}{\textbf{Fig.}}
    \caption{ \textbf{Quantum critical cooling and \rev{diverging} Gr\"uneisen ratio $\Gamma_g$.}
    \textbf{a} The isentropic lines of 1D quantum Ising model near QCP are illustrated in the $g$-$T$ plane 
    with $h=0$. Blue dashed lines represent the crossover lines enclosing the QCR. 
    \textbf{b} The Gr\"uneisen ratio $\Gamma_g$ for each fixed temperature. 
    The inset collects the peak/dip locations $(\tilde{g}_{\rm cr}, \rev{T})$, which exhibit a power-law 
    scaling \rev{$\tilde{g}_{\rm cr}\propto T^{1/z\nu}$}. 
    }
    \label{figm1}
    \end{figure}

To capture the pronounced low-temperature cooling effect clearly visible in the isentropic lines of Fig.~\ref{figm1}\textbf{a}, \rev{we propose that the peak/dip values of $\Gamma_g$ exhibit the characteristic $T^{-1/z\nu}$ scaling,} even near such a self-dual QCP. From the scaling form of the free energy, the Gr\"uneisen ratio satisfies $\Gamma_g = T^{-1/z\nu} \phi_{\Gamma_g}(y)$, where $\phi_{\Gamma_g}(y)$ with $y\equiv \tilde{g} \cdot T^{-1/z\nu}$ is the scaling function. The extrema of isothermal $\Gamma_g$ correspond to those of $\phi_{\Gamma_g}(y)$, since $\partial \Gamma_g /\partial g = T^{-2/z\nu} \phi_{\Gamma_g}'(y) = 0$ reduces to $\phi_{\Gamma_g}'(y)=0$. Consequently, each extremum of $\phi_{\Gamma_g}(y)$ generates a series of corresponding extrema in isothermal $\Gamma_g$, which naturally define the crossover lines. For instance, considering the peak\rev{/dip} condition $y=\rev{\pm}y_0$, the peak locations follow universal scaling \rev{$g_{\rm cr}=\pm y_0 T^{1/z\nu}\propto T^{1/z\nu}$} (see inset of Fig.~\ref{figm1}\textbf{b}), while the peak values satisfy $\Gamma_g = T^{-1/z\nu} \phi_{\Gamma_g}(\rev{\pm}y_0) \propto T^{-1/z\nu}$ [with $z\nu=1$ for the (1+1)D Ising universality class] as shown in Fig.~\ref{fig2}\textbf{e} in the main text. 
\\

\noindent
\textbf{Supercritical scaling analysis of $\Gamma_h$.}
Here, we conduct the scaling analysis of the quantum supercritical MCE induced by longitudinal fields. In the vicinity of a QCP, the singular part of entropy has a universal form $S = T^{{d}/{z}} \psi_S \left( h T^{-\Delta/{z\nu}}, \tilde{g} T^{-1/{z\nu}} \right)$~\cite{continentino2017}. For $\tilde{g}=0$, the hyperscaling function reduces to a single-variable scaling function $\psi_S(x,0) \rightarrow \phi_S(x)$, with $x \equiv h \cdot T^{-\Delta/{z\nu}}$. Moreover, the isothermal entropy derivative $(\partial S/\partial h)_T=T^{d/z-\Delta/z\nu}\phi'_S(x)$ is parity odd and the specific heat $C=T(\partial S/\partial T)_h=(d/z) \, T^{d/z}[\phi_S(x)-x\phi'_S(x)\Delta/d\nu]$ is parity even. \rev{As Gr\"uneisen ratio $\Gamma_h \equiv \frac{1}{T}\left(\frac{\partial T}{\partial h}\right)_S=-\frac{(\partial S/\partial h)_T}{T(\partial S/\partial T)_h}$}~\cite{Markus2005}, it thus also possesses the universal form $\Gamma_h=T^{-\Delta/z\nu}\phi_{\Gamma_h}(x)$, with the scaling function $\phi_{\Gamma_h}(x)\equiv - (z/d)\phi_S'(x)/[\phi_S(x)-x\phi'_S(x)\Delta/d\nu]$ \rev{(see Fig.~\ref{fig2}\textbf{f})}. \rev{The peak/dip positions of $\phi_{\Gamma_h}(x)$ at $x = \pm x_0$ define the quantum supercritical crossover lines $h_{\rm cr} = \pm x_0 T^{\Delta/z\nu} \propto T^{\Delta/z\nu}$, where the peak/dip values satisfy $\Gamma_h = T^{-\Delta/z\nu} \phi_{\Gamma_h}(\pm x_0) \propto T^{-\Delta/z\nu}$.} In the limit $x \ll 1$, we can further simplify the expression of $\phi_{\Gamma_h}(x)$. As $\phi_S(x)$ is a parity-even smooth function, we expand $\phi_S(x) = \sum_{i=0}^{\infty}A_{2i} x^{2i}$ for small $x$ and arrive at 
\begin{equation}
\begin{split}
\Gamma_h = \rev{-\frac{(\partial S/\partial h)_T}{T(\partial S/\partial T)_h}} \approx - \frac{T^{{d}/{z}-\Delta/{z\nu}} 2A_2 x}{({d}/{z})T^{{d}/{z}}A_0} \sim T^{-\Delta/{z\nu}} x.
\end{split}
\label{Gammah}
\end{equation}
This is consistent with the results in Fig.~2\textbf{f} of the main text, where $\Gamma_h$ has a universal form $\Gamma_h=T^{-\Delta/{z\nu}} \phi_{\Gamma_h}(x)$, with $\phi_{\Gamma_h}(x) \propto x$ for small $x$. Furthermore, when fixing the longitudinal field $h$ and varying temperature $T$, from Eq.~(\ref{Gammah}) we find a scaling law $\Gamma_h/h \propto T^{-2\Delta/{z\nu}}$ (see numerical results in the Supplementary Note \rev{3}).
\\

\noindent
\textbf{Heisenberg-Dzyaloshinskii-Moriya (HDM) model.}
Here, we introduce the HDM system that also hosts the quantum supercritical phenomena. The Hamiltonian of HDM model is 
$H = J \sum_{i} \textit{\textbf{S}}_i \cdot \textit{\textbf{S}}_{i+1} + \sum_{i} (-1)^i D_z \hat{z} \cdot (\textit{\textbf{S}}_i \times \textit{\textbf{S}}_{i+1}) - B \sum_i S_i^y$, where the DM interaction $D_z$ lies along the $z$-direction and the magnetic field $B$ is applied perpendicular to $z$-direction. Different unitary transformations can be implemented to the two sublattices as $D_z$ is alternating~\cite{Oshikawa1997Gaps}, and it results in an easy-plane XXZ model 
\begin{equation} 
    \begin{split}
H/J &= \frac{1}{\lambda} \sum_i \left( S_{i}^{x}S_{i+1}^{x} + S_{i}^{y}S_{i+1}^{y} 
+ \lambda S_{i}^{z} S_{i+1}^{z} \right) \notag \\
&\quad - g \sum_i S_{i}^{y} - h_s \sum_{i} (-1)^{i} S_i^{x}.
    \end{split}
    \notag
\end{equation}
Here, a uniform field $g = (B/J)\cos{\alpha}$ and a staggered field $h_s = ({B}/{J})\sin{\alpha}$ are involved, where $\alpha = \frac{1}{2} {\rm atan}{(D_z/J)}$ and the easy-plane anisotropy $\lambda = {1}/{\sqrt{1+(D_z/J)^2}} < 1$. The simulated MCE results of the HDM chain can be found in Supplementary Note 5.

In the easy-plane XXZ model, the field $g$ controls quantum fluctuations, and a Berezinskii-Kosterlitz-Thouless (BKT) quantum phase transition occurs at $g=0$; while $h_s$ breaks the $\mathbb{Z} _2$ symmetry and drives quantum supercriticality. According to the sine-Gordon model, the QSR crossovers follow $T\propto h_s^{1/(2-p)}$~\cite{Oshikawa1997Gaps, fradkin2013field}, where $1/4 ~ \rev{{\rm (XY~limit)}} \leq p  \leq 1/2 ~ \rev{{\rm (Heisenberg~limit)}}$ is the scaling dimension. Note that as the exponent $\Delta$ is not defined for the BKT transition, its supercritical exponent $\tilde{p}\equiv 2-p$ is peculiar for the 1D HDM chain considered here. In Fig.~\ref{fig4}\textbf{b} of the main text, we present the results of $\Gamma_{h_s} \equiv \frac{1}{T} (\frac{\partial T}{\partial h_s})_S \propto 1/T^{\tilde{p}}$, where a quantum supercritical scaling $\Gamma_{h_s}\propto T^{-3/2}$ is observed for small $D_z\rev{=0.1}$ \rev{($\lambda\simeq 0.995$ close to the Heisenberg limit)}. 
\\

\noindent
\textbf{Critical exponents of typical universality classes.}
The existence of universality classes with unique critical exponents is a fundamental aspect of QCPs not limited to Ising systems. \rev{A thorough examination of common universality classes, as listed in Table~\ref{Tab:Exponents}, reveals that $\Delta > 1$ consistently holds.} This ensures that the MCE of QSR is always significantly enhanced compared to that of QCR. In Tab.~\ref{Tab:Exponents}, for the conventional $(d+1)$D universality classes at dimensions $d\leq 2$ (below the upper critical dimension), scaling relations yield $\Delta/z\nu = (d+3-\eta)/2$. This results in quantum supercritical exponents of $(2-\eta/2)$ for $d=1$ and $(5/2-\eta/2)$ for $d=2$.  Given that $\eta/2$ is usually a small positive value, this implies $\Delta/z\nu \rev{\approx} (d+3)/2$. Consequently, we expect the Gr\"uneisen ratio $\Gamma_h$ to exhibit a significantly stronger divergence in 2D lattices than in 1D chains, \rev{e.g., $\Delta/z\nu \simeq 2.482$ (2D) versus $1.875$ (1D) for Ising systems}. Furthermore, our numerical simulations can capture this distinct scaling behavior, as detailed in Supplementary Note \rev{2}.

\begin{table}[h]
    \centering
    \begin{tabular}{lcccccc}
    \hline
    \textbf{Univ. Class} & $\beta$ & $\gamma$ & $\Delta$ & $z\nu$ & $\Delta/z\nu$ & Refs. \\ 
    \hline
    (1+1)D Ising & 1/8 & 7/4 & 15/8 & 1 & 15/8 & \cite{Henkel1999} \\ 
    (2+1)D Ising & 0.326 & 1.237 & 1.564 & 0.630 & 2.482 & \cite{3DIsing2025} \\ 
    (2+1)D XY & 0.349 & 1.318 & 1.667 & 0.672 & 2.481 & \cite{Deng2019} \\
    (2+1)D O(3) & 0.369 & 1.396 & 1.765 & 0.711 & 2.482 & \cite{3DHeisenberg2002} \\
    (1+1)D 3-Potts & 1/9 & 13/9 & 14/9 & 5/6 & 28/15 & \cite{Henkel1999} \\
    (1+1)D 4-Potts & 1/12 & 7/6 & 5/4 & 2/3 & 15/8 & \cite{Henkel1999} \\
    Mean Field & 1/2 & 1 & 3/2 & 1 & 3/2 & \cite{sachdev2015} \\
    \hline
    \end{tabular}
    \caption{Critical exponents of common universality classes.}
    \label{Tab:Exponents}
\end{table}

\noindent
\textbf{Data availability}
Source data are provided in this paper. The data generated in this study have been deposited in the Zenodo database [\href{https://doi.org/10.5281/zenodo.17243253}{https://doi.org/10.5281/zenodo.17243253}]. \rev{All other data that support the findings of this study are available from the corresponding authors upon request.} \rev{The experimental data in Fig.~4b are adapted from Refs.~\cite{Wolf2014,Oliver2017}, and the critical exponents in Tab.~1 are from Refs.~\cite{Henkel1999,3DHeisenberg2002,3DIsing2025,sachdev2015,Deng2019}.}\\

\noindent
\textbf{Code availability}
Numerical codes are provided in this paper. The relevant codes in this study have been deposited in the Zenodo database [\href{https://doi.org/10.5281/zenodo.16919271}{https://doi.org/10.5281/zenodo.16919271}]. 
\\

\bibliography{qsc}

\begin{thebibliography}{85}%
\makeatletter
\providecommand \@ifxundefined [1]{%
 \@ifx{#1\undefined}
}%
\providecommand \@ifnum [1]{%
 \ifnum #1\expandafter \@firstoftwo
 \else \expandafter \@secondoftwo
 \fi
}%
\providecommand \@ifx [1]{%
 \ifx #1\expandafter \@firstoftwo
 \else \expandafter \@secondoftwo
 \fi
}%
\providecommand \natexlab [1]{#1}%
\providecommand \enquote  [1]{``#1''}%
\providecommand \bibnamefont  [1]{#1}%
\providecommand \bibfnamefont [1]{#1}%
\providecommand \citenamefont [1]{#1}%
\providecommand \href@noop [0]{\@secondoftwo}%
\providecommand \href [0]{\begingroup \@sanitize@url \@href}%
\providecommand \@href[1]{\@@startlink{#1}\@@href}%
\providecommand \@@href[1]{\endgroup#1\@@endlink}%
\providecommand \@sanitize@url [0]{\catcode `\\12\catcode `\$12\catcode
  `\&12\catcode `\#12\catcode `\^12\catcode `\_12\catcode `\%12\relax}%
\providecommand \@@startlink[1]{}%
\providecommand \@@endlink[0]{}%
\providecommand \url  [0]{\begingroup\@sanitize@url \@url }%
\providecommand \@url [1]{\endgroup\@href {#1}{\urlprefix }}%
\providecommand \urlprefix  [0]{URL }%
\providecommand \Eprint [0]{\href }%
\providecommand \doibase [0]{https://doi.org/}%
\providecommand \selectlanguage [0]{\@gobble}%
\providecommand \bibinfo  [0]{\@secondoftwo}%
\providecommand \bibfield  [0]{\@secondoftwo}%
\providecommand \translation [1]{[#1]}%
\providecommand \BibitemOpen [0]{}%
\providecommand \bibitemStop [0]{}%
\providecommand \bibitemNoStop [0]{.\EOS\space}%
\providecommand \EOS [0]{\spacefactor3000\relax}%
\providecommand \BibitemShut  [1]{\csname bibitem#1\endcsname}%
\let\auto@bib@innerbib\@empty
\bibitem [{\citenamefont {Sachdev}(2000)}]{Sachdev2000}%
  \BibitemOpen
  \bibfield  {author} {\bibinfo {author} {\bibfnamefont {S.}~\bibnamefont
  {Sachdev}},\ }\bibfield  {title} {\bibinfo {title} {Quantum criticality:
  Competing ground states in low dimensions},\ }\href
  {https://doi.org/10.1126/science.288.5465.475} {\bibfield  {journal}
  {\bibinfo  {journal} {Science}\ }\textbf {\bibinfo {volume} {288}},\ \bibinfo
  {pages} {475} (\bibinfo {year} {2000})}\BibitemShut {NoStop}%
\bibitem [{\citenamefont {Coleman}\ and\ \citenamefont
  {Schofield}(2005)}]{Coleman2005}%
  \BibitemOpen
  \bibfield  {author} {\bibinfo {author} {\bibfnamefont {P.}~\bibnamefont
  {Coleman}}\ and\ \bibinfo {author} {\bibfnamefont {A.~J.}\ \bibnamefont
  {Schofield}},\ }\bibfield  {title} {\bibinfo {title} {Quantum criticality},\
  }\href {https://doi.org/https://doi.org/10.1038/nature03279} {\bibfield
  {journal} {\bibinfo  {journal} {Nature}\ }\textbf {\bibinfo {volume} {433}},\
  \bibinfo {pages} {226} (\bibinfo {year} {2005})}\BibitemShut {NoStop}%
\bibitem [{\citenamefont {Sachdev}(2015)}]{sachdev2015}%
  \BibitemOpen
  \bibfield  {author} {\bibinfo {author} {\bibfnamefont {S.}~\bibnamefont
  {Sachdev}},\ }\href@noop {} {\emph {\bibinfo {title} {Quantum phase
  transitions}}},\ \bibinfo {edition} {2nd}\ ed.\ (\bibinfo  {publisher}
  {Cambridge University Press},\ \bibinfo {address} {Cambridge},\ \bibinfo
  {year} {2015})\BibitemShut {NoStop}%
\bibitem [{\citenamefont {Sondhi}\ \emph {et~al.}(1997)\citenamefont {Sondhi},
  \citenamefont {Girvin}, \citenamefont {Carini},\ and\ \citenamefont
  {Shahar}}]{sondhi1997}%
  \BibitemOpen
  \bibfield  {author} {\bibinfo {author} {\bibfnamefont {S.~L.}\ \bibnamefont
  {Sondhi}}, \bibinfo {author} {\bibfnamefont {S.~M.}\ \bibnamefont {Girvin}},
  \bibinfo {author} {\bibfnamefont {J.~P.}\ \bibnamefont {Carini}},\ and\
  \bibinfo {author} {\bibfnamefont {D.}~\bibnamefont {Shahar}},\ }\bibfield
  {title} {\bibinfo {title} {Continuous quantum phase transitions},\ }\href
  {https://doi.org/10.1103/revmodphys.69.315} {\bibfield  {journal} {\bibinfo
  {journal} {Rev. Mod. Phys.}\ }\textbf {\bibinfo {volume} {69}},\ \bibinfo
  {pages} {315} (\bibinfo {year} {1997})}\BibitemShut {NoStop}%
\bibitem [{\citenamefont {Zhang}\ \emph {et~al.}(2012)\citenamefont {Zhang},
  \citenamefont {Hung}, \citenamefont {Tung},\ and\ \citenamefont
  {Chin}}]{Xibo2012}%
  \BibitemOpen
  \bibfield  {author} {\bibinfo {author} {\bibfnamefont {X.}~\bibnamefont
  {Zhang}}, \bibinfo {author} {\bibfnamefont {C.-L.}\ \bibnamefont {Hung}},
  \bibinfo {author} {\bibfnamefont {S.-K.}\ \bibnamefont {Tung}},\ and\
  \bibinfo {author} {\bibfnamefont {C.}~\bibnamefont {Chin}},\ }\bibfield
  {title} {\bibinfo {title} {Observation of quantum criticality with ultracold
  atoms in optical lattices},\ }\href {https://doi.org/10.1126/science.1217990}
  {\bibfield  {journal} {\bibinfo  {journal} {Science}\ }\textbf {\bibinfo
  {volume} {335}},\ \bibinfo {pages} {1070} (\bibinfo {year}
  {2012})}\BibitemShut {NoStop}%
\bibitem [{\citenamefont {Sachdev}\ and\ \citenamefont
  {Müller}(2009)}]{Sachdev2009}%
  \BibitemOpen
  \bibfield  {author} {\bibinfo {author} {\bibfnamefont {S.}~\bibnamefont
  {Sachdev}}\ and\ \bibinfo {author} {\bibfnamefont {M.}~\bibnamefont
  {Müller}},\ }\bibfield  {title} {\bibinfo {title} {Quantum criticality and
  black holes},\ }\href {https://doi.org/10.1088/0953-8984/21/16/164216}
  {\bibfield  {journal} {\bibinfo  {journal} {J. Phys.: Condens. Matter}\
  }\textbf {\bibinfo {volume} {21}},\ \bibinfo {pages} {164216} (\bibinfo
  {year} {2009})}\BibitemShut {NoStop}%
\bibitem [{\citenamefont {Čubrović}\ \emph {et~al.}(2009)\citenamefont
  {Čubrović}, \citenamefont {Zaanen},\ and\ \citenamefont
  {Schalm}}]{Mihailo2009}%
  \BibitemOpen
  \bibfield  {author} {\bibinfo {author} {\bibfnamefont {M.}~\bibnamefont
  {Čubrović}}, \bibinfo {author} {\bibfnamefont {J.}~\bibnamefont {Zaanen}},\
  and\ \bibinfo {author} {\bibfnamefont {K.}~\bibnamefont {Schalm}},\
  }\bibfield  {title} {\bibinfo {title} {String theory, quantum phase
  transitions, and the emergent {F}ermi liquid},\ }\href
  {https://doi.org/10.1126/science.1174962} {\bibfield  {journal} {\bibinfo
  {journal} {Science}\ }\textbf {\bibinfo {volume} {325}},\ \bibinfo {pages}
  {439} (\bibinfo {year} {2009})}\BibitemShut {NoStop}%
\bibitem [{\citenamefont {Sachdev}\ and\ \citenamefont
  {Ye}(1992)}]{Sachdev1992}%
  \BibitemOpen
  \bibfield  {author} {\bibinfo {author} {\bibfnamefont {S.}~\bibnamefont
  {Sachdev}}\ and\ \bibinfo {author} {\bibfnamefont {J.}~\bibnamefont {Ye}},\
  }\bibfield  {title} {\bibinfo {title} {Universal quantum-critical dynamics of
  two-dimensional antiferromagnets},\ }\href
  {https://doi.org/10.1103/PhysRevLett.69.2411} {\bibfield  {journal} {\bibinfo
   {journal} {Phys. Rev. Lett.}\ }\textbf {\bibinfo {volume} {69}},\ \bibinfo
  {pages} {2411} (\bibinfo {year} {1992})}\BibitemShut {NoStop}%
\bibitem [{\citenamefont {Kirchner}\ \emph {et~al.}(2020)\citenamefont
  {Kirchner}, \citenamefont {Paschen}, \citenamefont {Chen}, \citenamefont
  {Wirth}, \citenamefont {Feng}, \citenamefont {Thompson},\ and\ \citenamefont
  {Si}}]{Krichner2020}%
  \BibitemOpen
  \bibfield  {author} {\bibinfo {author} {\bibfnamefont {S.}~\bibnamefont
  {Kirchner}}, \bibinfo {author} {\bibfnamefont {S.}~\bibnamefont {Paschen}},
  \bibinfo {author} {\bibfnamefont {Q.}~\bibnamefont {Chen}}, \bibinfo {author}
  {\bibfnamefont {S.}~\bibnamefont {Wirth}}, \bibinfo {author} {\bibfnamefont
  {D.}~\bibnamefont {Feng}}, \bibinfo {author} {\bibfnamefont {J.~D.}\
  \bibnamefont {Thompson}},\ and\ \bibinfo {author} {\bibfnamefont
  {Q.}~\bibnamefont {Si}},\ }\bibfield  {title} {\bibinfo {title} {Colloquium:
  Heavy-electron quantum criticality and single-particle spectroscopy},\ }\href
  {https://doi.org/10.1103/RevModPhys.92.011002} {\bibfield  {journal}
  {\bibinfo  {journal} {Rev. Mod. Phys.}\ }\textbf {\bibinfo {volume} {92}},\
  \bibinfo {pages} {011002} (\bibinfo {year} {2020})}\BibitemShut {NoStop}%
\bibitem [{\citenamefont {Continentino}(2017)}]{continentino2017}%
  \BibitemOpen
  \bibfield  {author} {\bibinfo {author} {\bibfnamefont {M.}~\bibnamefont
  {Continentino}},\ }\href@noop {} {\emph {\bibinfo {title} {Quantum scaling in
  many-body systems}}},\ \bibinfo {edition} {2nd}\ ed.\ (\bibinfo  {publisher}
  {Cambridge University Press},\ \bibinfo {address} {Cambridge},\ \bibinfo
  {year} {2017})\BibitemShut {NoStop}%
\bibitem [{\citenamefont {Coldea}\ \emph {et~al.}(2010)\citenamefont {Coldea},
  \citenamefont {Tennant}, \citenamefont {Wheeler}, \citenamefont {Wawrzynska},
  \citenamefont {Prabhakaran}, \citenamefont {Telling}, \citenamefont
  {Habicht}, \citenamefont {Smeibidl},\ and\ \citenamefont
  {Kiefer}}]{coldea2010}%
  \BibitemOpen
  \bibfield  {author} {\bibinfo {author} {\bibfnamefont {R.}~\bibnamefont
  {Coldea}}, \bibinfo {author} {\bibfnamefont {D.~A.}\ \bibnamefont {Tennant}},
  \bibinfo {author} {\bibfnamefont {E.~M.}\ \bibnamefont {Wheeler}}, \bibinfo
  {author} {\bibfnamefont {E.}~\bibnamefont {Wawrzynska}}, \bibinfo {author}
  {\bibfnamefont {D.}~\bibnamefont {Prabhakaran}}, \bibinfo {author}
  {\bibfnamefont {M.}~\bibnamefont {Telling}}, \bibinfo {author} {\bibfnamefont
  {K.}~\bibnamefont {Habicht}}, \bibinfo {author} {\bibfnamefont
  {P.}~\bibnamefont {Smeibidl}},\ and\ \bibinfo {author} {\bibfnamefont
  {K.}~\bibnamefont {Kiefer}},\ }\bibfield  {title} {\bibinfo {title} {Quantum
  criticality in an {Ising} chain: Experimental evidence for emergent ${E}_8$
  symmetry},\ }\href {https://doi.org/10.1126/science.1180085} {\bibfield
  {journal} {\bibinfo  {journal} {Science}\ }\textbf {\bibinfo {volume}
  {327}},\ \bibinfo {pages} {177} (\bibinfo {year} {2010})}\BibitemShut
  {NoStop}%
\bibitem [{\citenamefont {Kinross}\ \emph {et~al.}(2014)\citenamefont
  {Kinross}, \citenamefont {Fu}, \citenamefont {Munsie}, \citenamefont
  {Dabkowska}, \citenamefont {Luke}, \citenamefont {Sachdev},\ and\
  \citenamefont {Imai}}]{kinross2014}%
  \BibitemOpen
  \bibfield  {author} {\bibinfo {author} {\bibfnamefont {A.~W.}\ \bibnamefont
  {Kinross}}, \bibinfo {author} {\bibfnamefont {M.}~\bibnamefont {Fu}},
  \bibinfo {author} {\bibfnamefont {T.~J.}\ \bibnamefont {Munsie}}, \bibinfo
  {author} {\bibfnamefont {H.~A.}\ \bibnamefont {Dabkowska}}, \bibinfo {author}
  {\bibfnamefont {G.~M.}\ \bibnamefont {Luke}}, \bibinfo {author}
  {\bibfnamefont {S.}~\bibnamefont {Sachdev}},\ and\ \bibinfo {author}
  {\bibfnamefont {T.}~\bibnamefont {Imai}},\ }\bibfield  {title} {\bibinfo
  {title} {Evolution of quantum fluctuations near the quantum critical point of
  the transverse field {I}sing chain system \ch{CoNb_2O_6}},\ }\href
  {https://doi.org/10.1103/PhysRevX.4.031008} {\bibfield  {journal} {\bibinfo
  {journal} {Phys. Rev. X}\ }\textbf {\bibinfo {volume} {4}},\ \bibinfo {pages}
  {031008} (\bibinfo {year} {2014})}\BibitemShut {NoStop}%
\bibitem [{\citenamefont {Wu}\ \emph {et~al.}(2014)\citenamefont {Wu},
  \citenamefont {Kormos},\ and\ \citenamefont {Si}}]{wu2014}%
  \BibitemOpen
  \bibfield  {author} {\bibinfo {author} {\bibfnamefont {J.}~\bibnamefont
  {Wu}}, \bibinfo {author} {\bibfnamefont {M.}~\bibnamefont {Kormos}},\ and\
  \bibinfo {author} {\bibfnamefont {Q.}~\bibnamefont {Si}},\ }\bibfield
  {title} {\bibinfo {title} {Finite-temperature spin dynamics in a perturbed
  quantum critical {I}sing chain with an ${E}_{8}$ symmetry},\ }\href
  {https://doi.org/10.1103/PhysRevLett.113.247201} {\bibfield  {journal}
  {\bibinfo  {journal} {Phys. Rev. Lett.}\ }\textbf {\bibinfo {volume} {113}},\
  \bibinfo {pages} {247201} (\bibinfo {year} {2014})}\BibitemShut {NoStop}%
\bibitem [{\citenamefont {Morris}\ \emph {et~al.}(2014)\citenamefont {Morris},
  \citenamefont {Vald\'es~Aguilar}, \citenamefont {Ghosh}, \citenamefont
  {Koohpayeh}, \citenamefont {Krizan}, \citenamefont {Cava}, \citenamefont
  {Tchernyshyov}, \citenamefont {McQueen},\ and\ \citenamefont
  {Armitage}}]{morris2014}%
  \BibitemOpen
  \bibfield  {author} {\bibinfo {author} {\bibfnamefont {C.~M.}\ \bibnamefont
  {Morris}}, \bibinfo {author} {\bibfnamefont {R.}~\bibnamefont
  {Vald\'es~Aguilar}}, \bibinfo {author} {\bibfnamefont {A.}~\bibnamefont
  {Ghosh}}, \bibinfo {author} {\bibfnamefont {S.~M.}\ \bibnamefont
  {Koohpayeh}}, \bibinfo {author} {\bibfnamefont {J.}~\bibnamefont {Krizan}},
  \bibinfo {author} {\bibfnamefont {R.~J.}\ \bibnamefont {Cava}}, \bibinfo
  {author} {\bibfnamefont {O.}~\bibnamefont {Tchernyshyov}}, \bibinfo {author}
  {\bibfnamefont {T.~M.}\ \bibnamefont {McQueen}},\ and\ \bibinfo {author}
  {\bibfnamefont {N.~P.}\ \bibnamefont {Armitage}},\ }\bibfield  {title}
  {\bibinfo {title} {Hierarchy of bound states in the one-dimensional
  ferromagnetic {I}sing chain \ch{CoNb_2O_6} investigated by high-resolution
  time-domain terahertz spectroscopy},\ }\href
  {https://doi.org/10.1103/PhysRevLett.112.137403} {\bibfield  {journal}
  {\bibinfo  {journal} {Phys. Rev. Lett.}\ }\textbf {\bibinfo {volume} {112}},\
  \bibinfo {pages} {137403} (\bibinfo {year} {2014})}\BibitemShut {NoStop}%
\bibitem [{\citenamefont {Liang}\ \emph {et~al.}(2015)\citenamefont {Liang},
  \citenamefont {Koohpayeh}, \citenamefont {Krizan}, \citenamefont {McQueen},
  \citenamefont {Cava},\ and\ \citenamefont {Ong}}]{liang2015}%
  \BibitemOpen
  \bibfield  {author} {\bibinfo {author} {\bibfnamefont {T.}~\bibnamefont
  {Liang}}, \bibinfo {author} {\bibfnamefont {S.~M.}\ \bibnamefont
  {Koohpayeh}}, \bibinfo {author} {\bibfnamefont {J.~W.}\ \bibnamefont
  {Krizan}}, \bibinfo {author} {\bibfnamefont {T.~M.}\ \bibnamefont {McQueen}},
  \bibinfo {author} {\bibfnamefont {R.~J.}\ \bibnamefont {Cava}},\ and\
  \bibinfo {author} {\bibfnamefont {N.~P.}\ \bibnamefont {Ong}},\ }\bibfield
  {title} {\bibinfo {title} {Heat capacity peak at the quantum critical point
  of the transverse {I}sing magnet \ch{CoNb_2O_6}},\ }\href
  {https://doi.org/10.1038/ncomms8611} {\bibfield  {journal} {\bibinfo
  {journal} {Nat. Commun.}\ }\textbf {\bibinfo {volume} {6}},\ \bibinfo {pages}
  {7611} (\bibinfo {year} {2015})}\BibitemShut {NoStop}%
\bibitem [{\citenamefont {Fava}\ \emph {et~al.}(2020)\citenamefont {Fava},
  \citenamefont {Coldea},\ and\ \citenamefont {Parameswaran}}]{Fava2020}%
  \BibitemOpen
  \bibfield  {author} {\bibinfo {author} {\bibfnamefont {M.}~\bibnamefont
  {Fava}}, \bibinfo {author} {\bibfnamefont {R.}~\bibnamefont {Coldea}},\ and\
  \bibinfo {author} {\bibfnamefont {S.~A.}\ \bibnamefont {Parameswaran}},\
  }\bibfield  {title} {\bibinfo {title} {Glide symmetry breaking and {I}sing
  criticality in the quasi-1{D} magnet \ch{CoNb_2O_6}},\ }\href
  {https://doi.org/10.1073/pnas.2007986117} {\bibfield  {journal} {\bibinfo
  {journal} {Proc. Natl. Acad. Sci.}\ }\textbf {\bibinfo {volume} {117}},\
  \bibinfo {pages} {25219} (\bibinfo {year} {2020})}\BibitemShut {NoStop}%
\bibitem [{\citenamefont {Amelin}\ \emph {et~al.}(2020)\citenamefont {Amelin},
  \citenamefont {Engelmayer}, \citenamefont {Viirok}, \citenamefont {Nagel},
  \citenamefont {R{\~o}{\~o}m}, \citenamefont {Lorenz},\ and\ \citenamefont
  {Wang}}]{amelin2020}%
  \BibitemOpen
  \bibfield  {author} {\bibinfo {author} {\bibfnamefont {K.}~\bibnamefont
  {Amelin}}, \bibinfo {author} {\bibfnamefont {J.}~\bibnamefont {Engelmayer}},
  \bibinfo {author} {\bibfnamefont {J.}~\bibnamefont {Viirok}}, \bibinfo
  {author} {\bibfnamefont {U.}~\bibnamefont {Nagel}}, \bibinfo {author}
  {\bibfnamefont {T.}~\bibnamefont {R{\~o}{\~o}m}}, \bibinfo {author}
  {\bibfnamefont {T.}~\bibnamefont {Lorenz}},\ and\ \bibinfo {author}
  {\bibfnamefont {Z.}~\bibnamefont {Wang}},\ }\bibfield  {title} {\bibinfo
  {title} {Experimental observation of quantum many-body excitations of
  ${E}_{8}$ symmetry in the {I}sing chain ferromagnet \ch{CoNb_2O_6}},\ }\href
  {https://doi.org/10.1103/PhysRevB.102.104431} {\bibfield  {journal} {\bibinfo
   {journal} {Phys. Rev. B}\ }\textbf {\bibinfo {volume} {102}},\ \bibinfo
  {pages} {104431} (\bibinfo {year} {2020})}\BibitemShut {NoStop}%
\bibitem [{\citenamefont {Morris}\ \emph {et~al.}(2021)\citenamefont {Morris},
  \citenamefont {Desai}, \citenamefont {Viirok}, \citenamefont {H{\"u}vonen},
  \citenamefont {Nagel}, \citenamefont {R{\~o}{\~o}m}, \citenamefont {Krizan},
  \citenamefont {Cava}, \citenamefont {McQueen}, \citenamefont {Koohpayeh},
  \citenamefont {Kaul},\ and\ \citenamefont {Armitage}}]{morris2021}%
  \BibitemOpen
  \bibfield  {author} {\bibinfo {author} {\bibfnamefont {C.~M.}\ \bibnamefont
  {Morris}}, \bibinfo {author} {\bibfnamefont {N.}~\bibnamefont {Desai}},
  \bibinfo {author} {\bibfnamefont {J.}~\bibnamefont {Viirok}}, \bibinfo
  {author} {\bibfnamefont {D.}~\bibnamefont {H{\"u}vonen}}, \bibinfo {author}
  {\bibfnamefont {U.}~\bibnamefont {Nagel}}, \bibinfo {author} {\bibfnamefont
  {T.}~\bibnamefont {R{\~o}{\~o}m}}, \bibinfo {author} {\bibfnamefont {J.~W.}\
  \bibnamefont {Krizan}}, \bibinfo {author} {\bibfnamefont {R.~J.}\
  \bibnamefont {Cava}}, \bibinfo {author} {\bibfnamefont {T.~M.}\ \bibnamefont
  {McQueen}}, \bibinfo {author} {\bibfnamefont {S.~M.}\ \bibnamefont
  {Koohpayeh}}, \bibinfo {author} {\bibfnamefont {R.~K.}\ \bibnamefont
  {Kaul}},\ and\ \bibinfo {author} {\bibfnamefont {N.~P.}\ \bibnamefont
  {Armitage}},\ }\bibfield  {title} {\bibinfo {title} {Duality and domain wall
  dynamics in a twisted {K}itaev chain},\ }\href
  {https://doi.org/10.1038/s41567-021-01208-0} {\bibfield  {journal} {\bibinfo
  {journal} {Nat. Phys.}\ }\textbf {\bibinfo {volume} {17}},\ \bibinfo {pages}
  {832} (\bibinfo {year} {2021})}\BibitemShut {NoStop}%
\bibitem [{\citenamefont {Xu}\ \emph {et~al.}(2022)\citenamefont {Xu},
  \citenamefont {Wang}, \citenamefont {Huang}, \citenamefont {Ni},
  \citenamefont {Zhao}, \citenamefont {Dai}, \citenamefont {Pan}, \citenamefont
  {Hong}, \citenamefont {Chauhan}, \citenamefont {Koohpayeh}, \citenamefont
  {Armitage},\ and\ \citenamefont {Li}}]{xu2022}%
  \BibitemOpen
  \bibfield  {author} {\bibinfo {author} {\bibfnamefont {Y.}~\bibnamefont
  {Xu}}, \bibinfo {author} {\bibfnamefont {L.~S.}\ \bibnamefont {Wang}},
  \bibinfo {author} {\bibfnamefont {Y.~Y.}\ \bibnamefont {Huang}}, \bibinfo
  {author} {\bibfnamefont {J.~M.}\ \bibnamefont {Ni}}, \bibinfo {author}
  {\bibfnamefont {C.~C.}\ \bibnamefont {Zhao}}, \bibinfo {author}
  {\bibfnamefont {Y.~F.}\ \bibnamefont {Dai}}, \bibinfo {author} {\bibfnamefont
  {B.~Y.}\ \bibnamefont {Pan}}, \bibinfo {author} {\bibfnamefont {X.~C.}\
  \bibnamefont {Hong}}, \bibinfo {author} {\bibfnamefont {P.}~\bibnamefont
  {Chauhan}}, \bibinfo {author} {\bibfnamefont {S.~M.}\ \bibnamefont
  {Koohpayeh}}, \bibinfo {author} {\bibfnamefont {N.~P.}\ \bibnamefont
  {Armitage}},\ and\ \bibinfo {author} {\bibfnamefont {S.~Y.}\ \bibnamefont
  {Li}},\ }\bibfield  {title} {\bibinfo {title} {Quantum critical magnetic
  excitations in spin-$1/2$ and spin-1 chain systems},\ }\href
  {https://doi.org/10.1103/PhysRevX.12.021020} {\bibfield  {journal} {\bibinfo
  {journal} {Phys. Rev. X}\ }\textbf {\bibinfo {volume} {12}},\ \bibinfo
  {pages} {021020} (\bibinfo {year} {2022})}\BibitemShut {NoStop}%
\bibitem [{\citenamefont {Woodland}\ \emph {et~al.}(2023)\citenamefont
  {Woodland}, \citenamefont {Macdougal}, \citenamefont {Cabrera}, \citenamefont
  {Thompson}, \citenamefont {Prabhakaran}, \citenamefont {Bewley},\ and\
  \citenamefont {Coldea}}]{woodland2023}%
  \BibitemOpen
  \bibfield  {author} {\bibinfo {author} {\bibfnamefont {L.}~\bibnamefont
  {Woodland}}, \bibinfo {author} {\bibfnamefont {D.}~\bibnamefont {Macdougal}},
  \bibinfo {author} {\bibfnamefont {I.~M.}\ \bibnamefont {Cabrera}}, \bibinfo
  {author} {\bibfnamefont {J.~D.}\ \bibnamefont {Thompson}}, \bibinfo {author}
  {\bibfnamefont {D.}~\bibnamefont {Prabhakaran}}, \bibinfo {author}
  {\bibfnamefont {R.~I.}\ \bibnamefont {Bewley}},\ and\ \bibinfo {author}
  {\bibfnamefont {R.}~\bibnamefont {Coldea}},\ }\bibfield  {title} {\bibinfo
  {title} {Tuning the confinement potential between spinons in the {I}sing
  chain \ch{CoNb_2O_6} using longitudinal fields and quantitative determination
  of the microscopic {H}amiltonian},\ }\href
  {https://doi.org/10.1103/physrevb.108.184416} {\bibfield  {journal} {\bibinfo
   {journal} {Phys. Rev. B}\ }\textbf {\bibinfo {volume} {108}},\ \bibinfo
  {pages} {184416} (\bibinfo {year} {2023})}\BibitemShut {NoStop}%
\bibitem [{\citenamefont {Xi}\ \emph {et~al.}(2024{\natexlab{a}})\citenamefont
  {Xi}, \citenamefont {Wang}, \citenamefont {Gao}, \citenamefont {Jiang},
  \citenamefont {Yu},\ and\ \citenamefont {Wu}}]{Ning2024}%
  \BibitemOpen
  \bibfield  {author} {\bibinfo {author} {\bibfnamefont {N.}~\bibnamefont
  {Xi}}, \bibinfo {author} {\bibfnamefont {X.}~\bibnamefont {Wang}}, \bibinfo
  {author} {\bibfnamefont {Y.}~\bibnamefont {Gao}}, \bibinfo {author}
  {\bibfnamefont {Y.}~\bibnamefont {Jiang}}, \bibinfo {author} {\bibfnamefont
  {R.}~\bibnamefont {Yu}},\ and\ \bibinfo {author} {\bibfnamefont
  {J.}~\bibnamefont {Wu}},\ }\href@noop {} {\bibinfo {title} {Emergent
  ${D}_8^{(1)}$ spectrum and topological soliton excitation in \ch{CoNb_2O_6}}}
  (\bibinfo {year} {2024}{\natexlab{a}}),\ \Eprint
  {https://arxiv.org/abs/2403.10785} {arXiv:2403.10785} \BibitemShut {NoStop}%
\bibitem [{\citenamefont {Kimura}\ \emph {et~al.}(2007)\citenamefont {Kimura},
  \citenamefont {Yashiro}, \citenamefont {Okunishi}, \citenamefont {Hagiwara},
  \citenamefont {He}, \citenamefont {Kindo}, \citenamefont {Taniyama},\ and\
  \citenamefont {Itoh}}]{Kimura2007}%
  \BibitemOpen
  \bibfield  {author} {\bibinfo {author} {\bibfnamefont {S.}~\bibnamefont
  {Kimura}}, \bibinfo {author} {\bibfnamefont {H.}~\bibnamefont {Yashiro}},
  \bibinfo {author} {\bibfnamefont {K.}~\bibnamefont {Okunishi}}, \bibinfo
  {author} {\bibfnamefont {M.}~\bibnamefont {Hagiwara}}, \bibinfo {author}
  {\bibfnamefont {Z.}~\bibnamefont {He}}, \bibinfo {author} {\bibfnamefont
  {K.}~\bibnamefont {Kindo}}, \bibinfo {author} {\bibfnamefont
  {T.}~\bibnamefont {Taniyama}},\ and\ \bibinfo {author} {\bibfnamefont
  {M.}~\bibnamefont {Itoh}},\ }\bibfield  {title} {\bibinfo {title}
  {Field-induced order-disorder transition in antiferromagnetic
  \ch{BaCo_2V_2O_8} driven by a softening of spinon excitation},\ }\href
  {https://doi.org/10.1103/PhysRevLett.99.087602} {\bibfield  {journal}
  {\bibinfo  {journal} {Phys. Rev. Lett.}\ }\textbf {\bibinfo {volume} {99}},\
  \bibinfo {pages} {087602} (\bibinfo {year} {2007})}\BibitemShut {NoStop}%
\bibitem [{\citenamefont {Faure}\ \emph {et~al.}(2018)\citenamefont {Faure},
  \citenamefont {Takayoshi}, \citenamefont {Petit}, \citenamefont {Simonet},
  \citenamefont {Raymond}, \citenamefont {Regnault}, \citenamefont {Boehm},
  \citenamefont {White}, \citenamefont {Månsson}, \citenamefont {Rüegg},
  \citenamefont {Lejay}, \citenamefont {Canals}, \citenamefont {Lorenz},
  \citenamefont {Furuya}, \citenamefont {Giamarchi},\ and\ \citenamefont
  {Grenier}}]{faure2018}%
  \BibitemOpen
  \bibfield  {author} {\bibinfo {author} {\bibfnamefont {Q.}~\bibnamefont
  {Faure}}, \bibinfo {author} {\bibfnamefont {S.}~\bibnamefont {Takayoshi}},
  \bibinfo {author} {\bibfnamefont {S.}~\bibnamefont {Petit}}, \bibinfo
  {author} {\bibfnamefont {V.}~\bibnamefont {Simonet}}, \bibinfo {author}
  {\bibfnamefont {S.}~\bibnamefont {Raymond}}, \bibinfo {author} {\bibfnamefont
  {L.-P.}\ \bibnamefont {Regnault}}, \bibinfo {author} {\bibfnamefont
  {M.}~\bibnamefont {Boehm}}, \bibinfo {author} {\bibfnamefont {J.~S.}\
  \bibnamefont {White}}, \bibinfo {author} {\bibfnamefont {M.}~\bibnamefont
  {Månsson}}, \bibinfo {author} {\bibfnamefont {C.}~\bibnamefont {Rüegg}},
  \bibinfo {author} {\bibfnamefont {P.}~\bibnamefont {Lejay}}, \bibinfo
  {author} {\bibfnamefont {B.}~\bibnamefont {Canals}}, \bibinfo {author}
  {\bibfnamefont {T.}~\bibnamefont {Lorenz}}, \bibinfo {author} {\bibfnamefont
  {S.~C.}\ \bibnamefont {Furuya}}, \bibinfo {author} {\bibfnamefont
  {T.}~\bibnamefont {Giamarchi}},\ and\ \bibinfo {author} {\bibfnamefont
  {B.}~\bibnamefont {Grenier}},\ }\bibfield  {title} {\bibinfo {title}
  {Topological quantum phase transition in the {I}sing-like antiferromagnetic
  spin chain \ch{BaCo_2V_2O_8}},\ }\href
  {https://doi.org/10.1038/s41567-018-0126-8} {\bibfield  {journal} {\bibinfo
  {journal} {Nat. Phys.}\ }\textbf {\bibinfo {volume} {14}},\ \bibinfo {pages}
  {716} (\bibinfo {year} {2018})}\BibitemShut {NoStop}%
\bibitem [{\citenamefont {Wang}\ \emph
  {et~al.}(2018{\natexlab{a}})\citenamefont {Wang}, \citenamefont {Lorenz},
  \citenamefont {Gorbunov}, \citenamefont {Cong}, \citenamefont {Kohama},
  \citenamefont {Niesen}, \citenamefont {Breunig}, \citenamefont {Engelmayer},
  \citenamefont {Herman}, \citenamefont {Wu}, \citenamefont {Kindo},
  \citenamefont {Wosnitza}, \citenamefont {Zherlitsyn},\ and\ \citenamefont
  {Loidl}}]{wang2018}%
  \BibitemOpen
  \bibfield  {author} {\bibinfo {author} {\bibfnamefont {Z.}~\bibnamefont
  {Wang}}, \bibinfo {author} {\bibfnamefont {T.}~\bibnamefont {Lorenz}},
  \bibinfo {author} {\bibfnamefont {D.~I.}\ \bibnamefont {Gorbunov}}, \bibinfo
  {author} {\bibfnamefont {P.~T.}\ \bibnamefont {Cong}}, \bibinfo {author}
  {\bibfnamefont {Y.}~\bibnamefont {Kohama}}, \bibinfo {author} {\bibfnamefont
  {S.}~\bibnamefont {Niesen}}, \bibinfo {author} {\bibfnamefont
  {O.}~\bibnamefont {Breunig}}, \bibinfo {author} {\bibfnamefont
  {J.}~\bibnamefont {Engelmayer}}, \bibinfo {author} {\bibfnamefont
  {A.}~\bibnamefont {Herman}}, \bibinfo {author} {\bibfnamefont
  {J.}~\bibnamefont {Wu}}, \bibinfo {author} {\bibfnamefont {K.}~\bibnamefont
  {Kindo}}, \bibinfo {author} {\bibfnamefont {J.}~\bibnamefont {Wosnitza}},
  \bibinfo {author} {\bibfnamefont {S.}~\bibnamefont {Zherlitsyn}},\ and\
  \bibinfo {author} {\bibfnamefont {A.}~\bibnamefont {Loidl}},\ }\bibfield
  {title} {\bibinfo {title} {Quantum criticality of an {Ising-like} spin-$1/2$
  antiferromagnetic chain in a transverse magnetic field},\ }\href
  {https://doi.org/10.1103/PhysRevLett.120.207205} {\bibfield  {journal}
  {\bibinfo  {journal} {Phys. Rev. Lett.}\ }\textbf {\bibinfo {volume} {120}},\
  \bibinfo {pages} {207205} (\bibinfo {year} {2018}{\natexlab{a}})}\BibitemShut
  {NoStop}%
\bibitem [{\citenamefont {Wang}\ \emph {et~al.}(2019)\citenamefont {Wang},
  \citenamefont {Schmidt}, \citenamefont {Loidl}, \citenamefont {Wu},
  \citenamefont {Zou}, \citenamefont {Yang}, \citenamefont {Dong},
  \citenamefont {Kohama}, \citenamefont {Kindo}, \citenamefont {Gorbunov},
  \citenamefont {Niesen}, \citenamefont {Breunig}, \citenamefont {Engelmayer},\
  and\ \citenamefont {Lorenz}}]{wang2019}%
  \BibitemOpen
  \bibfield  {author} {\bibinfo {author} {\bibfnamefont {Z.}~\bibnamefont
  {Wang}}, \bibinfo {author} {\bibfnamefont {M.}~\bibnamefont {Schmidt}},
  \bibinfo {author} {\bibfnamefont {A.}~\bibnamefont {Loidl}}, \bibinfo
  {author} {\bibfnamefont {J.}~\bibnamefont {Wu}}, \bibinfo {author}
  {\bibfnamefont {H.}~\bibnamefont {Zou}}, \bibinfo {author} {\bibfnamefont
  {W.}~\bibnamefont {Yang}}, \bibinfo {author} {\bibfnamefont {C.}~\bibnamefont
  {Dong}}, \bibinfo {author} {\bibfnamefont {Y.}~\bibnamefont {Kohama}},
  \bibinfo {author} {\bibfnamefont {K.}~\bibnamefont {Kindo}}, \bibinfo
  {author} {\bibfnamefont {D.~I.}\ \bibnamefont {Gorbunov}}, \bibinfo {author}
  {\bibfnamefont {S.}~\bibnamefont {Niesen}}, \bibinfo {author} {\bibfnamefont
  {O.}~\bibnamefont {Breunig}}, \bibinfo {author} {\bibfnamefont
  {J.}~\bibnamefont {Engelmayer}},\ and\ \bibinfo {author} {\bibfnamefont
  {T.}~\bibnamefont {Lorenz}},\ }\bibfield  {title} {\bibinfo {title} {Quantum
  critical dynamics of a {Heisenberg-Ising} chain in a longitudinal field:
  Many-body strings versus fractional excitations},\ }\href
  {https://doi.org/10.1103/PhysRevLett.123.067202} {\bibfield  {journal}
  {\bibinfo  {journal} {Phys. Rev. Lett.}\ }\textbf {\bibinfo {volume} {123}},\
  \bibinfo {pages} {067202} (\bibinfo {year} {2019})}\BibitemShut {NoStop}%
\bibitem [{\citenamefont {Zou}\ \emph {et~al.}(2021)\citenamefont {Zou},
  \citenamefont {Cui}, \citenamefont {Wang}, \citenamefont {Zhang},
  \citenamefont {Yang}, \citenamefont {Xu}, \citenamefont {Okutani},
  \citenamefont {Hagiwara}, \citenamefont {Matsuda}, \citenamefont {Wang},
  \citenamefont {Mussardo}, \citenamefont {H\'ods\'agi}, \citenamefont
  {Kormos}, \citenamefont {He}, \citenamefont {Kimura}, \citenamefont {Yu},
  \citenamefont {Yu}, \citenamefont {Ma},\ and\ \citenamefont {Wu}}]{zou2021}%
  \BibitemOpen
  \bibfield  {author} {\bibinfo {author} {\bibfnamefont {H.}~\bibnamefont
  {Zou}}, \bibinfo {author} {\bibfnamefont {Y.}~\bibnamefont {Cui}}, \bibinfo
  {author} {\bibfnamefont {X.}~\bibnamefont {Wang}}, \bibinfo {author}
  {\bibfnamefont {Z.}~\bibnamefont {Zhang}}, \bibinfo {author} {\bibfnamefont
  {J.}~\bibnamefont {Yang}}, \bibinfo {author} {\bibfnamefont {G.}~\bibnamefont
  {Xu}}, \bibinfo {author} {\bibfnamefont {A.}~\bibnamefont {Okutani}},
  \bibinfo {author} {\bibfnamefont {M.}~\bibnamefont {Hagiwara}}, \bibinfo
  {author} {\bibfnamefont {M.}~\bibnamefont {Matsuda}}, \bibinfo {author}
  {\bibfnamefont {G.}~\bibnamefont {Wang}}, \bibinfo {author} {\bibfnamefont
  {G.}~\bibnamefont {Mussardo}}, \bibinfo {author} {\bibfnamefont
  {K.}~\bibnamefont {H\'ods\'agi}}, \bibinfo {author} {\bibfnamefont
  {M.}~\bibnamefont {Kormos}}, \bibinfo {author} {\bibfnamefont
  {Z.}~\bibnamefont {He}}, \bibinfo {author} {\bibfnamefont {S.}~\bibnamefont
  {Kimura}}, \bibinfo {author} {\bibfnamefont {R.}~\bibnamefont {Yu}}, \bibinfo
  {author} {\bibfnamefont {W.}~\bibnamefont {Yu}}, \bibinfo {author}
  {\bibfnamefont {J.}~\bibnamefont {Ma}},\ and\ \bibinfo {author}
  {\bibfnamefont {J.}~\bibnamefont {Wu}},\ }\bibfield  {title} {\bibinfo
  {title} {${E}_{8}$ spectra of quasi-one-dimensional antiferromagnet
  \ch{BaCo_2V_2O_8} under transverse field},\ }\href
  {https://doi.org/10.1103/PhysRevLett.127.077201} {\bibfield  {journal}
  {\bibinfo  {journal} {Phys. Rev. Lett.}\ }\textbf {\bibinfo {volume} {127}},\
  \bibinfo {pages} {077201} (\bibinfo {year} {2021})}\BibitemShut {NoStop}%
\bibitem [{\citenamefont {Wang}\ \emph
  {et~al.}(2024{\natexlab{a}})\citenamefont {Wang}, \citenamefont {Halati},
  \citenamefont {Bernier}, \citenamefont {Ponomaryov}, \citenamefont
  {Gorbunov}, \citenamefont {Niesen}, \citenamefont {Breunig}, \citenamefont
  {Klopf}, \citenamefont {Zvyagin}, \citenamefont {Lorenz}, \citenamefont
  {Loidl},\ and\ \citenamefont {Kollath}}]{Wang2024Nature}%
  \BibitemOpen
  \bibfield  {author} {\bibinfo {author} {\bibfnamefont {Z.}~\bibnamefont
  {Wang}}, \bibinfo {author} {\bibfnamefont {C.-M.}\ \bibnamefont {Halati}},
  \bibinfo {author} {\bibfnamefont {J.-S.}\ \bibnamefont {Bernier}}, \bibinfo
  {author} {\bibfnamefont {A.}~\bibnamefont {Ponomaryov}}, \bibinfo {author}
  {\bibfnamefont {D.~I.}\ \bibnamefont {Gorbunov}}, \bibinfo {author}
  {\bibfnamefont {S.}~\bibnamefont {Niesen}}, \bibinfo {author} {\bibfnamefont
  {O.}~\bibnamefont {Breunig}}, \bibinfo {author} {\bibfnamefont {J.~M.}\
  \bibnamefont {Klopf}}, \bibinfo {author} {\bibfnamefont {S.}~\bibnamefont
  {Zvyagin}}, \bibinfo {author} {\bibfnamefont {T.}~\bibnamefont {Lorenz}},
  \bibinfo {author} {\bibfnamefont {A.}~\bibnamefont {Loidl}},\ and\ \bibinfo
  {author} {\bibfnamefont {C.}~\bibnamefont {Kollath}},\ }\bibfield  {title}
  {\bibinfo {title} {Experimental observation of repulsively bound magnons},\
  }\href {https://doi.org/10.1038/s41586-024-07599-3} {\bibfield  {journal}
  {\bibinfo  {journal} {Nature}\ }\textbf {\bibinfo {volume} {631}},\ \bibinfo
  {pages} {760} (\bibinfo {year} {2024}{\natexlab{a}})}\BibitemShut {NoStop}%
\bibitem [{\citenamefont {He}\ \emph {et~al.}(2006)\citenamefont {He},
  \citenamefont {Taniyama},\ and\ \citenamefont {Itoh}}]{He2006}%
  \BibitemOpen
  \bibfield  {author} {\bibinfo {author} {\bibfnamefont {Z.}~\bibnamefont
  {He}}, \bibinfo {author} {\bibfnamefont {T.}~\bibnamefont {Taniyama}},\ and\
  \bibinfo {author} {\bibfnamefont {M.}~\bibnamefont {Itoh}},\ }\bibfield
  {title} {\bibinfo {title} {Antiferromagnetic-paramagnetic transitions in
  longitudinal and transverse magnetic fields in a \ch{SrCo_2V_2O_8} crystal},\
  }\href {https://doi.org/10.1103/PhysRevB.73.212406} {\bibfield  {journal}
  {\bibinfo  {journal} {Phys. Rev. B}\ }\textbf {\bibinfo {volume} {73}},\
  \bibinfo {pages} {212406} (\bibinfo {year} {2006})}\BibitemShut {NoStop}%
\bibitem [{\citenamefont {Wang}\ \emph {et~al.}(2015)\citenamefont {Wang},
  \citenamefont {Schmidt}, \citenamefont {Bera}, \citenamefont {Islam},
  \citenamefont {Lake}, \citenamefont {Loidl},\ and\ \citenamefont
  {Deisenhofer}}]{Wang2015}%
  \BibitemOpen
  \bibfield  {author} {\bibinfo {author} {\bibfnamefont {Z.}~\bibnamefont
  {Wang}}, \bibinfo {author} {\bibfnamefont {M.}~\bibnamefont {Schmidt}},
  \bibinfo {author} {\bibfnamefont {A.~K.}\ \bibnamefont {Bera}}, \bibinfo
  {author} {\bibfnamefont {A.~T. M.~N.}\ \bibnamefont {Islam}}, \bibinfo
  {author} {\bibfnamefont {B.}~\bibnamefont {Lake}}, \bibinfo {author}
  {\bibfnamefont {A.}~\bibnamefont {Loidl}},\ and\ \bibinfo {author}
  {\bibfnamefont {J.}~\bibnamefont {Deisenhofer}},\ }\bibfield  {title}
  {\bibinfo {title} {Spinon confinement in the one-dimensional {I}sing-like
  antiferromagnet \ch{SrCo_2V_2O_8}},\ }\href
  {https://doi.org/10.1103/PhysRevB.91.140404} {\bibfield  {journal} {\bibinfo
  {journal} {Phys. Rev. B}\ }\textbf {\bibinfo {volume} {91}},\ \bibinfo
  {pages} {140404} (\bibinfo {year} {2015})}\BibitemShut {NoStop}%
\bibitem [{\citenamefont {Bera}\ \emph {et~al.}(2017)\citenamefont {Bera},
  \citenamefont {Lake}, \citenamefont {Essler}, \citenamefont {Vanderstraeten},
  \citenamefont {Hubig}, \citenamefont {Schollw\"ock}, \citenamefont {Islam},
  \citenamefont {Schneidewind},\ and\ \citenamefont
  {Quintero-Castro}}]{Bera2017}%
  \BibitemOpen
  \bibfield  {author} {\bibinfo {author} {\bibfnamefont {A.~K.}\ \bibnamefont
  {Bera}}, \bibinfo {author} {\bibfnamefont {B.}~\bibnamefont {Lake}}, \bibinfo
  {author} {\bibfnamefont {F.~H.~L.}\ \bibnamefont {Essler}}, \bibinfo {author}
  {\bibfnamefont {L.}~\bibnamefont {Vanderstraeten}}, \bibinfo {author}
  {\bibfnamefont {C.}~\bibnamefont {Hubig}}, \bibinfo {author} {\bibfnamefont
  {U.}~\bibnamefont {Schollw\"ock}}, \bibinfo {author} {\bibfnamefont {A.~T.
  M.~N.}\ \bibnamefont {Islam}}, \bibinfo {author} {\bibfnamefont
  {A.}~\bibnamefont {Schneidewind}},\ and\ \bibinfo {author} {\bibfnamefont
  {D.~L.}\ \bibnamefont {Quintero-Castro}},\ }\bibfield  {title} {\bibinfo
  {title} {Spinon confinement in a quasi-one-dimensional anisotropic
  {H}eisenberg magnet},\ }\href {https://doi.org/10.1103/PhysRevB.96.054423}
  {\bibfield  {journal} {\bibinfo  {journal} {Phys. Rev. B}\ }\textbf {\bibinfo
  {volume} {96}},\ \bibinfo {pages} {054423} (\bibinfo {year}
  {2017})}\BibitemShut {NoStop}%
\bibitem [{\citenamefont {Cui}\ \emph {et~al.}(2019)\citenamefont {Cui},
  \citenamefont {Zou}, \citenamefont {Xi}, \citenamefont {He}, \citenamefont
  {Yang}, \citenamefont {Shu}, \citenamefont {Zhang}, \citenamefont {Hu},
  \citenamefont {Chen}, \citenamefont {Yu}, \citenamefont {Wu},\ and\
  \citenamefont {Yu}}]{cui2019}%
  \BibitemOpen
  \bibfield  {author} {\bibinfo {author} {\bibfnamefont {Y.}~\bibnamefont
  {Cui}}, \bibinfo {author} {\bibfnamefont {H.}~\bibnamefont {Zou}}, \bibinfo
  {author} {\bibfnamefont {N.}~\bibnamefont {Xi}}, \bibinfo {author}
  {\bibfnamefont {Z.}~\bibnamefont {He}}, \bibinfo {author} {\bibfnamefont
  {Y.~X.}\ \bibnamefont {Yang}}, \bibinfo {author} {\bibfnamefont
  {L.}~\bibnamefont {Shu}}, \bibinfo {author} {\bibfnamefont {G.~H.}\
  \bibnamefont {Zhang}}, \bibinfo {author} {\bibfnamefont {Z.}~\bibnamefont
  {Hu}}, \bibinfo {author} {\bibfnamefont {T.}~\bibnamefont {Chen}}, \bibinfo
  {author} {\bibfnamefont {R.}~\bibnamefont {Yu}}, \bibinfo {author}
  {\bibfnamefont {J.}~\bibnamefont {Wu}},\ and\ \bibinfo {author}
  {\bibfnamefont {W.}~\bibnamefont {Yu}},\ }\bibfield  {title} {\bibinfo
  {title} {Quantum criticality of the {I}sing-like screw chain antiferromagnet
  \ch{SrCo_2V_2O_8} in a transverse magnetic field},\ }\href
  {https://doi.org/10.1103/PhysRevLett.123.067203} {\bibfield  {journal}
  {\bibinfo  {journal} {Phys. Rev. Lett.}\ }\textbf {\bibinfo {volume} {123}},\
  \bibinfo {pages} {067203} (\bibinfo {year} {2019})}\BibitemShut {NoStop}%
\bibitem [{\citenamefont {Wang}\ \emph
  {et~al.}(2018{\natexlab{b}})\citenamefont {Wang}, \citenamefont {Halati},
  \citenamefont {Bernier}, \citenamefont {Ponomaryov}, \citenamefont
  {Gorbunov}, \citenamefont {Niesen}, \citenamefont {Breunig}, \citenamefont
  {Klopf}, \citenamefont {Zvyagin}, \citenamefont {Lorenz}, \citenamefont
  {Loidl},\ and\ \citenamefont {Kollath}}]{Wang2018Nature}%
  \BibitemOpen
  \bibfield  {author} {\bibinfo {author} {\bibfnamefont {Z.}~\bibnamefont
  {Wang}}, \bibinfo {author} {\bibfnamefont {C.-M.}\ \bibnamefont {Halati}},
  \bibinfo {author} {\bibfnamefont {J.-S.}\ \bibnamefont {Bernier}}, \bibinfo
  {author} {\bibfnamefont {A.}~\bibnamefont {Ponomaryov}}, \bibinfo {author}
  {\bibfnamefont {D.~I.}\ \bibnamefont {Gorbunov}}, \bibinfo {author}
  {\bibfnamefont {S.}~\bibnamefont {Niesen}}, \bibinfo {author} {\bibfnamefont
  {O.}~\bibnamefont {Breunig}}, \bibinfo {author} {\bibfnamefont {J.~M.}\
  \bibnamefont {Klopf}}, \bibinfo {author} {\bibfnamefont {S.}~\bibnamefont
  {Zvyagin}}, \bibinfo {author} {\bibfnamefont {T.}~\bibnamefont {Lorenz}},
  \bibinfo {author} {\bibfnamefont {A.}~\bibnamefont {Loidl}},\ and\ \bibinfo
  {author} {\bibfnamefont {C.}~\bibnamefont {Kollath}},\ }\bibfield  {title}
  {\bibinfo {title} {Experimental observation of {B}ethe strings},\ }\href
  {https://doi.org/10.1038/nature25466} {\bibfield  {journal} {\bibinfo
  {journal} {Nature}\ }\textbf {\bibinfo {volume} {554}},\ \bibinfo {pages}
  {219} (\bibinfo {year} {2018}{\natexlab{b}})}\BibitemShut {NoStop}%
\bibitem [{\citenamefont {Zhu}\ \emph {et~al.}(2003)\citenamefont {Zhu},
  \citenamefont {Garst}, \citenamefont {Rosch},\ and\ \citenamefont
  {Si}}]{zhu2003}%
  \BibitemOpen
  \bibfield  {author} {\bibinfo {author} {\bibfnamefont {L.}~\bibnamefont
  {Zhu}}, \bibinfo {author} {\bibfnamefont {M.}~\bibnamefont {Garst}}, \bibinfo
  {author} {\bibfnamefont {A.}~\bibnamefont {Rosch}},\ and\ \bibinfo {author}
  {\bibfnamefont {Q.}~\bibnamefont {Si}},\ }\bibfield  {title} {\bibinfo
  {title} {Universally diverging {G}rüneisen parameter and the magnetocaloric
  effect close to quantum critical points},\ }\href
  {https://doi.org/10.1103/physrevlett.91.066404} {\bibfield  {journal}
  {\bibinfo  {journal} {Phys. Rev. Lett.}\ }\textbf {\bibinfo {volume} {91}},\
  \bibinfo {pages} {066404} (\bibinfo {year} {2003})}\BibitemShut {NoStop}%
\bibitem [{\citenamefont {{Zhitomirsky}}\ and\ \citenamefont
  {{Honecker}}(2004)}]{Zhitomirsky2004}%
  \BibitemOpen
  \bibfield  {author} {\bibinfo {author} {\bibfnamefont {M.~E.}\ \bibnamefont
  {{Zhitomirsky}}}\ and\ \bibinfo {author} {\bibfnamefont {A.}~\bibnamefont
  {{Honecker}}},\ }\bibfield  {title} {\bibinfo {title} {Magnetocaloric effect
  in one-dimensional antiferromagnets},\ }\href
  {https://doi.org/10.1088/1742-5468/2004/07/P07012} {\bibfield  {journal}
  {\bibinfo  {journal} {J. Stat. Mech.: Theor. Exp.}\ }\textbf {\bibinfo
  {volume} {2004}},\ \bibinfo {pages} {07012} (\bibinfo {year}
  {2004})}\BibitemShut {NoStop}%
\bibitem [{\citenamefont {Garst}\ and\ \citenamefont
  {Rosch}(2005)}]{Markus2005}%
  \BibitemOpen
  \bibfield  {author} {\bibinfo {author} {\bibfnamefont {M.}~\bibnamefont
  {Garst}}\ and\ \bibinfo {author} {\bibfnamefont {A.}~\bibnamefont {Rosch}},\
  }\bibfield  {title} {\bibinfo {title} {Sign change of the {G}r\"uneisen
  parameter and magnetocaloric effect near quantum critical points},\ }\href
  {https://doi.org/10.1103/PhysRevB.72.205129} {\bibfield  {journal} {\bibinfo
  {journal} {Phys. Rev. B}\ }\textbf {\bibinfo {volume} {72}},\ \bibinfo
  {pages} {205129} (\bibinfo {year} {2005})}\BibitemShut {NoStop}%
\bibitem [{\citenamefont {Honecker}\ and\ \citenamefont
  {Wessel}(2009)}]{Honecker2009}%
  \BibitemOpen
  \bibfield  {author} {\bibinfo {author} {\bibfnamefont {A.}~\bibnamefont
  {Honecker}}\ and\ \bibinfo {author} {\bibfnamefont {S.}~\bibnamefont
  {Wessel}},\ }\bibfield  {title} {\bibinfo {title} {Magnetocaloric effect in
  quantum spin-s chains},\ }\href {http://dx.doi.org/10.5488/CMP.12.3.399}
  {\bibfield  {journal} {\bibinfo  {journal} {Condens. Matter Phys.}\ }\textbf
  {\bibinfo {volume} {12}},\ \bibinfo {pages} {399} (\bibinfo {year}
  {2009})}\BibitemShut {NoStop}%
\bibitem [{\citenamefont {Wu}\ \emph {et~al.}(2011)\citenamefont {Wu},
  \citenamefont {Zhu},\ and\ \citenamefont {Si}}]{Wu2011JPhCS}%
  \BibitemOpen
  \bibfield  {author} {\bibinfo {author} {\bibfnamefont {J.}~\bibnamefont
  {Wu}}, \bibinfo {author} {\bibfnamefont {L.}~\bibnamefont {Zhu}},\ and\
  \bibinfo {author} {\bibfnamefont {Q.}~\bibnamefont {Si}},\ }\bibfield
  {title} {\bibinfo {title} {Entropy accumulation near quantum critical points:
  effects beyond hyperscaling},\ }\href
  {https://doi.org/10.1088/1742-6596/273/1/012019} {\bibfield  {journal}
  {\bibinfo  {journal} {J. Phys.: Conf. Ser.}\ }\textbf {\bibinfo {volume}
  {273}},\ \bibinfo {pages} {012019} (\bibinfo {year} {2011})}\BibitemShut
  {NoStop}%
\bibitem [{\citenamefont {Zhang}(2019)}]{zhang2019}%
  \BibitemOpen
  \bibfield  {author} {\bibinfo {author} {\bibfnamefont {L.}~\bibnamefont
  {Zhang}},\ }\bibfield  {title} {\bibinfo {title} {Universal thermodynamic
  signature of self-dual quantum critical points},\ }\href
  {https://doi.org/10.1103/physrevlett.123.230601} {\bibfield  {journal}
  {\bibinfo  {journal} {Phys. Rev. Lett.}\ }\textbf {\bibinfo {volume} {123}},\
  \bibinfo {pages} {230601} (\bibinfo {year} {2019})}\BibitemShut {NoStop}%
\bibitem [{\citenamefont {Wolf}\ \emph {et~al.}(2011)\citenamefont {Wolf},
  \citenamefont {Tsui}, \citenamefont {Jaiswal-Nagar}, \citenamefont {Tutsch},
  \citenamefont {Honecker}, \citenamefont {Removi{\'c}-Langer}, \citenamefont
  {Hofmann}, \citenamefont {Prokofiev}, \citenamefont {Assmus}, \citenamefont
  {Donath},\ and\ \citenamefont {Lang}}]{Wolf2011}%
  \BibitemOpen
  \bibfield  {author} {\bibinfo {author} {\bibfnamefont {B.}~\bibnamefont
  {Wolf}}, \bibinfo {author} {\bibfnamefont {Y.}~\bibnamefont {Tsui}}, \bibinfo
  {author} {\bibfnamefont {D.}~\bibnamefont {Jaiswal-Nagar}}, \bibinfo {author}
  {\bibfnamefont {U.}~\bibnamefont {Tutsch}}, \bibinfo {author} {\bibfnamefont
  {A.}~\bibnamefont {Honecker}}, \bibinfo {author} {\bibfnamefont
  {K.}~\bibnamefont {Removi{\'c}-Langer}}, \bibinfo {author} {\bibfnamefont
  {G.}~\bibnamefont {Hofmann}}, \bibinfo {author} {\bibfnamefont
  {A.}~\bibnamefont {Prokofiev}}, \bibinfo {author} {\bibfnamefont
  {W.}~\bibnamefont {Assmus}}, \bibinfo {author} {\bibfnamefont
  {G.}~\bibnamefont {Donath}},\ and\ \bibinfo {author} {\bibfnamefont
  {M.}~\bibnamefont {Lang}},\ }\bibfield  {title} {\bibinfo {title}
  {Magnetocaloric effect and magnetic cooling near a field-induced
  quantum-critical point},\ }\href {https://doi.org/10.1073/pnas.1017047108}
  {\bibfield  {journal} {\bibinfo  {journal} {Proc. Natl. Acad. Sci. U.S.A.}\
  }\textbf {\bibinfo {volume} {108}},\ \bibinfo {pages} {6862} (\bibinfo {year}
  {2011})}\BibitemShut {NoStop}%
\bibitem [{\citenamefont {Wolf}\ \emph {et~al.}(2014)\citenamefont {Wolf},
  \citenamefont {Honecker}, \citenamefont {Hofstetter}, \citenamefont
  {Tutsch},\ and\ \citenamefont {Lang}}]{Wolf2014}%
  \BibitemOpen
  \bibfield  {author} {\bibinfo {author} {\bibfnamefont {B.}~\bibnamefont
  {Wolf}}, \bibinfo {author} {\bibfnamefont {A.}~\bibnamefont {Honecker}},
  \bibinfo {author} {\bibfnamefont {W.}~\bibnamefont {Hofstetter}}, \bibinfo
  {author} {\bibfnamefont {U.}~\bibnamefont {Tutsch}},\ and\ \bibinfo {author}
  {\bibfnamefont {M.}~\bibnamefont {Lang}},\ }\bibfield  {title} {\bibinfo
  {title} {Cooling through quantum criticality and many-body effects in
  condensed matter and cold gases},\ }\href
  {https://doi.org/10.1142/s0217979214300175} {\bibfield  {journal} {\bibinfo
  {journal} {Int. J. Mod. Phys. B}\ }\textbf {\bibinfo {volume} {28}},\
  \bibinfo {pages} {1430017} (\bibinfo {year} {2014})}\BibitemShut {NoStop}%
\bibitem [{\citenamefont {Tokiwa}\ \emph {et~al.}(2009)\citenamefont {Tokiwa},
  \citenamefont {Radu}, \citenamefont {Geibel}, \citenamefont {Steglich},\ and\
  \citenamefont {Gegenwart}}]{Tokiwa2009}%
  \BibitemOpen
  \bibfield  {author} {\bibinfo {author} {\bibfnamefont {Y.}~\bibnamefont
  {Tokiwa}}, \bibinfo {author} {\bibfnamefont {T.}~\bibnamefont {Radu}},
  \bibinfo {author} {\bibfnamefont {C.}~\bibnamefont {Geibel}}, \bibinfo
  {author} {\bibfnamefont {F.}~\bibnamefont {Steglich}},\ and\ \bibinfo
  {author} {\bibfnamefont {P.}~\bibnamefont {Gegenwart}},\ }\bibfield  {title}
  {\bibinfo {title} {{Divergence of the magnetic Gr\"uneisen ratio at the
  field-induced quantum critical point in {YbRh$_{2}$Si$_{2}$}}},\ }\href
  {https://doi.org/10.1103/PhysRevLett.102.066401} {\bibfield  {journal}
  {\bibinfo  {journal} {Phys. Rev. Lett.}\ }\textbf {\bibinfo {volume} {102}},\
  \bibinfo {pages} {066401} (\bibinfo {year} {2009})}\BibitemShut {NoStop}%
\bibitem [{\citenamefont {Tokiwa}\ \emph {et~al.}(2015)\citenamefont {Tokiwa},
  \citenamefont {Stingl}, \citenamefont {Kim}, \citenamefont {Takabatake},\
  and\ \citenamefont {Gegenwart}}]{Tokiwa2015Signature}%
  \BibitemOpen
  \bibfield  {author} {\bibinfo {author} {\bibfnamefont {Y.}~\bibnamefont
  {Tokiwa}}, \bibinfo {author} {\bibfnamefont {C.}~\bibnamefont {Stingl}},
  \bibinfo {author} {\bibfnamefont {M.-S.}\ \bibnamefont {Kim}}, \bibinfo
  {author} {\bibfnamefont {T.}~\bibnamefont {Takabatake}},\ and\ \bibinfo
  {author} {\bibfnamefont {P.}~\bibnamefont {Gegenwart}},\ }\bibfield  {title}
  {\bibinfo {title} {Characteristic signatures of quantum criticality driven by
  geometrical frustration},\ }\href {https://doi.org/10.1126/sciadv.1500001}
  {\bibfield  {journal} {\bibinfo  {journal} {Sci. Adv.}\ }\textbf {\bibinfo
  {volume} {1}},\ \bibinfo {pages} {e1500001} (\bibinfo {year}
  {2015})}\BibitemShut {NoStop}%
\bibitem [{\citenamefont {Gegenwart}(2016)}]{Gegenwart2016}%
  \BibitemOpen
  \bibfield  {author} {\bibinfo {author} {\bibfnamefont {P.}~\bibnamefont
  {Gegenwart}},\ }\bibfield  {title} {\bibinfo {title} {Gr{\"u}neisen parameter
  studies on heavy fermion quantum criticality},\ }\href
  {https://doi.org/10.1088/0034-4885/79/11/114502} {\bibfield  {journal}
  {\bibinfo  {journal} {Rep. Prog. Phys.}\ }\textbf {\bibinfo {volume} {79}},\
  \bibinfo {pages} {114502} (\bibinfo {year} {2016})}\BibitemShut {NoStop}%
\bibitem [{\citenamefont {Wolf}\ \emph {et~al.}(2016)\citenamefont {Wolf},
  \citenamefont {Tutsch}, \citenamefont {Dörschug}, \citenamefont {Krellner},
  \citenamefont {Ritter}, \citenamefont {Assmus},\ and\ \citenamefont
  {Lang}}]{Wolf2016}%
  \BibitemOpen
  \bibfield  {author} {\bibinfo {author} {\bibfnamefont {B.}~\bibnamefont
  {Wolf}}, \bibinfo {author} {\bibfnamefont {U.}~\bibnamefont {Tutsch}},
  \bibinfo {author} {\bibfnamefont {S.}~\bibnamefont {Dörschug}}, \bibinfo
  {author} {\bibfnamefont {C.}~\bibnamefont {Krellner}}, \bibinfo {author}
  {\bibfnamefont {F.}~\bibnamefont {Ritter}}, \bibinfo {author} {\bibfnamefont
  {W.}~\bibnamefont {Assmus}},\ and\ \bibinfo {author} {\bibfnamefont
  {M.}~\bibnamefont {Lang}},\ }\bibfield  {title} {\bibinfo {title} {Magnetic
  cooling close to a quantum phase transition—the case of
  {Er$_2$Ti$_2$O$_7$}},\ }\href {https://doi.org/10.1063/1.4961708} {\bibfield
  {journal} {\bibinfo  {journal} {J. Appl. Phys}\ }\textbf {\bibinfo {volume}
  {120}},\ \bibinfo {pages} {142112} (\bibinfo {year} {2016})}\BibitemShut
  {NoStop}%
\bibitem [{\citenamefont {Breunig}\ \emph {et~al.}(2017)\citenamefont
  {Breunig}, \citenamefont {Garst}, \citenamefont {Klümper}, \citenamefont
  {Rohrkamp}, \citenamefont {Turnbull},\ and\ \citenamefont
  {Lorenz}}]{Oliver2017}%
  \BibitemOpen
  \bibfield  {author} {\bibinfo {author} {\bibfnamefont {O.}~\bibnamefont
  {Breunig}}, \bibinfo {author} {\bibfnamefont {M.}~\bibnamefont {Garst}},
  \bibinfo {author} {\bibfnamefont {A.}~\bibnamefont {Klümper}}, \bibinfo
  {author} {\bibfnamefont {J.}~\bibnamefont {Rohrkamp}}, \bibinfo {author}
  {\bibfnamefont {M.~M.}\ \bibnamefont {Turnbull}},\ and\ \bibinfo {author}
  {\bibfnamefont {T.}~\bibnamefont {Lorenz}},\ }\bibfield  {title} {\bibinfo
  {title} {Quantum criticality in the spin-1/2 {H}eisenberg chain system copper
  pyrazine dinitrate},\ }\href {https://doi.org/10.1126/sciadv.aao3773}
  {\bibfield  {journal} {\bibinfo  {journal} {Sci. Adv.}\ }\textbf {\bibinfo
  {volume} {3}},\ \bibinfo {pages} {eaao3773} (\bibinfo {year}
  {2017})}\BibitemShut {NoStop}%
\bibitem [{\citenamefont {Xiang}\ \emph {et~al.}(2025)\citenamefont {Xiang},
  \citenamefont {Lv}, \citenamefont {Shen}, \citenamefont {Su}, \citenamefont
  {He}, \citenamefont {Zhu}, \citenamefont {Gao}, \citenamefont {Liu},
  \citenamefont {Qu}, \citenamefont {Wang}, \citenamefont {Chen}, \citenamefont
  {Zhao}, \citenamefont {Li}, \citenamefont {Li}, \citenamefont {Yang},
  \citenamefont {Luo}, \citenamefont {Sun}, \citenamefont {Jin}, \citenamefont
  {Qi}, \citenamefont {Zhou}, \citenamefont {Li},\ and\ \citenamefont
  {Su}}]{Xiang2025}%
  \BibitemOpen
  \bibfield  {author} {\bibinfo {author} {\bibfnamefont {J.}~\bibnamefont
  {Xiang}}, \bibinfo {author} {\bibfnamefont {E.}~\bibnamefont {Lv}}, \bibinfo
  {author} {\bibfnamefont {Q.}~\bibnamefont {Shen}}, \bibinfo {author}
  {\bibfnamefont {C.}~\bibnamefont {Su}}, \bibinfo {author} {\bibfnamefont
  {X.}~\bibnamefont {He}}, \bibinfo {author} {\bibfnamefont {Y.}~\bibnamefont
  {Zhu}}, \bibinfo {author} {\bibfnamefont {Y.}~\bibnamefont {Gao}}, \bibinfo
  {author} {\bibfnamefont {X.-Y.}\ \bibnamefont {Liu}}, \bibinfo {author}
  {\bibfnamefont {D.-W.}\ \bibnamefont {Qu}}, \bibinfo {author} {\bibfnamefont
  {X.}~\bibnamefont {Wang}}, \bibinfo {author} {\bibfnamefont {X.}~\bibnamefont
  {Chen}}, \bibinfo {author} {\bibfnamefont {Q.}~\bibnamefont {Zhao}}, \bibinfo
  {author} {\bibfnamefont {H.}~\bibnamefont {Li}}, \bibinfo {author}
  {\bibfnamefont {S.}~\bibnamefont {Li}}, \bibinfo {author} {\bibfnamefont
  {J.}~\bibnamefont {Yang}}, \bibinfo {author} {\bibfnamefont {J.}~\bibnamefont
  {Luo}}, \bibinfo {author} {\bibfnamefont {P.}~\bibnamefont {Sun}}, \bibinfo
  {author} {\bibfnamefont {W.}~\bibnamefont {Jin}}, \bibinfo {author}
  {\bibfnamefont {Y.}~\bibnamefont {Qi}}, \bibinfo {author} {\bibfnamefont
  {R.}~\bibnamefont {Zhou}}, \bibinfo {author} {\bibfnamefont {W.}~\bibnamefont
  {Li}},\ and\ \bibinfo {author} {\bibfnamefont {G.}~\bibnamefont {Su}},\
  }\href@noop {} {\bibinfo {title} {{Universal Magnetocaloric Effect near
  Quantum Critical Point of Magnon Bose-Einstein Condensation}}} (\bibinfo
  {year} {2025}),\ \Eprint {https://arxiv.org/abs/2508.05750}
  {arXiv:2508.05750} \BibitemShut {NoStop}%
\bibitem [{\citenamefont {Wang}\ \emph
  {et~al.}(2024{\natexlab{b}})\citenamefont {Wang}, \citenamefont {Lv},
  \citenamefont {Li}, \citenamefont {Jin},\ and\ \citenamefont
  {Li}}]{wang2024qsc}%
  \BibitemOpen
  \bibfield  {author} {\bibinfo {author} {\bibfnamefont {J.}~\bibnamefont
  {Wang}}, \bibinfo {author} {\bibfnamefont {E.}~\bibnamefont {Lv}}, \bibinfo
  {author} {\bibfnamefont {X.}~\bibnamefont {Li}}, \bibinfo {author}
  {\bibfnamefont {Y.}~\bibnamefont {Jin}},\ and\ \bibinfo {author}
  {\bibfnamefont {W.}~\bibnamefont {Li}},\ }\href@noop {} {\bibinfo {title}
  {Quantum supercritical crossover with dynamical singularity}} (\bibinfo
  {year} {2024}{\natexlab{b}}),\ \Eprint {https://arxiv.org/abs/2407.05455}
  {arXiv:2407.05455} \BibitemShut {NoStop}%
\bibitem [{\citenamefont {Cagniard de~la Tour}(1822)}]{Cagniard1822}%
  \BibitemOpen
  \bibfield  {author} {\bibinfo {author} {\bibfnamefont {C.}~\bibnamefont
  {Cagniard de~la Tour}},\ }\bibfield  {title} {\bibinfo {title} {Exposé de
  quelques résultats obtenu par l’action combinée de la chaleur et de la
  compression sur certains liquides, tels que l’eau, l’alcool, l’éther
  sulfurique et l’essence de pétrole rectifiée},\ }\href@noop {} {\bibfield
   {journal} {\bibinfo  {journal} {Ann. Chim. Phys.}\ }\textbf {\bibinfo
  {volume} {21}},\ \bibinfo {pages} {127} (\bibinfo {year} {1822})}\BibitemShut
  {NoStop}%
\bibitem [{\citenamefont {Andrews}(1869)}]{Andrews1869}%
  \BibitemOpen
  \bibfield  {author} {\bibinfo {author} {\bibfnamefont {T.}~\bibnamefont
  {Andrews}},\ }\bibfield  {title} {\bibinfo {title} {On the continuity of the
  gaseous and liquid states of matter},\ }\href@noop {} {\bibfield  {journal}
  {\bibinfo  {journal} {Philosophical Transactions of the Royal Society of
  London}\ }\textbf {\bibinfo {volume} {159}},\ \bibinfo {pages} {575}
  (\bibinfo {year} {1869})}\BibitemShut {NoStop}%
\bibitem [{\citenamefont {Xu}\ \emph {et~al.}(2005)\citenamefont {Xu},
  \citenamefont {Kumar}, \citenamefont {Buldyrev}, \citenamefont {Chen},
  \citenamefont {Poole}, \citenamefont {Sciortino},\ and\ \citenamefont
  {Stanley}}]{xu2005}%
  \BibitemOpen
  \bibfield  {author} {\bibinfo {author} {\bibfnamefont {L.}~\bibnamefont
  {Xu}}, \bibinfo {author} {\bibfnamefont {P.}~\bibnamefont {Kumar}}, \bibinfo
  {author} {\bibfnamefont {S.~V.}\ \bibnamefont {Buldyrev}}, \bibinfo {author}
  {\bibfnamefont {S.-H.}\ \bibnamefont {Chen}}, \bibinfo {author}
  {\bibfnamefont {P.~H.}\ \bibnamefont {Poole}}, \bibinfo {author}
  {\bibfnamefont {F.}~\bibnamefont {Sciortino}},\ and\ \bibinfo {author}
  {\bibfnamefont {H.~E.}\ \bibnamefont {Stanley}},\ }\bibfield  {title}
  {\bibinfo {title} {Relation between the {W}idom line and the dynamic
  crossover in systems with a liquid–liquid phase transition},\ }\href
  {https://doi.org/10.1073/pnas.0507870102} {\bibfield  {journal} {\bibinfo
  {journal} {Proc. Natl. Acad. Sci.}\ }\textbf {\bibinfo {volume} {102}},\
  \bibinfo {pages} {16558} (\bibinfo {year} {2005})}\BibitemShut {NoStop}%
\bibitem [{\citenamefont {Brazhkin}\ \emph {et~al.}(2013)\citenamefont
  {Brazhkin}, \citenamefont {Fomin}, \citenamefont {Lyapin}, \citenamefont
  {Ryzhov}, \citenamefont {Tsiok},\ and\ \citenamefont
  {Trachenko}}]{brazhkin2013}%
  \BibitemOpen
  \bibfield  {author} {\bibinfo {author} {\bibfnamefont {V.~V.}\ \bibnamefont
  {Brazhkin}}, \bibinfo {author} {\bibfnamefont {Y.~D.}\ \bibnamefont {Fomin}},
  \bibinfo {author} {\bibfnamefont {A.~G.}\ \bibnamefont {Lyapin}}, \bibinfo
  {author} {\bibfnamefont {V.~N.}\ \bibnamefont {Ryzhov}}, \bibinfo {author}
  {\bibfnamefont {E.~N.}\ \bibnamefont {Tsiok}},\ and\ \bibinfo {author}
  {\bibfnamefont {K.}~\bibnamefont {Trachenko}},\ }\bibfield  {title} {\bibinfo
  {title} {``liquid-gas'' transition in the supercritical region: Fundamental
  changes in the particle dynamics},\ }\href
  {https://doi.org/10.1103/PhysRevLett.111.145901} {\bibfield  {journal}
  {\bibinfo  {journal} {Phys. Rev. Lett.}\ }\textbf {\bibinfo {volume} {111}},\
  \bibinfo {pages} {145901} (\bibinfo {year} {2013})}\BibitemShut {NoStop}%
\bibitem [{\citenamefont {Li}\ and\ \citenamefont
  {Jin}(2024)}]{li2024supfluid}%
  \BibitemOpen
  \bibfield  {author} {\bibinfo {author} {\bibfnamefont {X.}~\bibnamefont
  {Li}}\ and\ \bibinfo {author} {\bibfnamefont {Y.}~\bibnamefont {Jin}},\
  }\bibfield  {title} {\bibinfo {title} {Thermodynamic crossovers in
  supercritical fluids},\ }\href {https://doi.org/10.1073/pnas.2400313121}
  {\bibfield  {journal} {\bibinfo  {journal} {Proc. Natl. Acad. Sci.}\ }\textbf
  {\bibinfo {volume} {121}} (\bibinfo {year} {2024})}\BibitemShut {NoStop}%
\bibitem [{\citenamefont {Li}\ \emph {et~al.}(2011)\citenamefont {Li},
  \citenamefont {Ran}, \citenamefont {Gong}, \citenamefont {Zhao},
  \citenamefont {Xi}, \citenamefont {Ye},\ and\ \citenamefont {Su}}]{li2011}%
  \BibitemOpen
  \bibfield  {author} {\bibinfo {author} {\bibfnamefont {W.}~\bibnamefont
  {Li}}, \bibinfo {author} {\bibfnamefont {S.-J.}\ \bibnamefont {Ran}},
  \bibinfo {author} {\bibfnamefont {S.-S.}\ \bibnamefont {Gong}}, \bibinfo
  {author} {\bibfnamefont {Y.}~\bibnamefont {Zhao}}, \bibinfo {author}
  {\bibfnamefont {B.}~\bibnamefont {Xi}}, \bibinfo {author} {\bibfnamefont
  {F.}~\bibnamefont {Ye}},\ and\ \bibinfo {author} {\bibfnamefont
  {G.}~\bibnamefont {Su}},\ }\bibfield  {title} {\bibinfo {title} {Linearized
  tensor renormalization group algorithm for the calculation of thermodynamic
  properties of quantum lattice models},\ }\href
  {https://doi.org/10.1103/PhysRevLett.106.127202} {\bibfield  {journal}
  {\bibinfo  {journal} {Phys. Rev. Lett.}\ }\textbf {\bibinfo {volume} {106}},\
  \bibinfo {pages} {127202} (\bibinfo {year} {2011})}\BibitemShut {NoStop}%
\bibitem [{\citenamefont {Dong}\ \emph {et~al.}(2017)\citenamefont {Dong},
  \citenamefont {Chen}, \citenamefont {Liu},\ and\ \citenamefont
  {Li}}]{Dong2017bilayer}%
  \BibitemOpen
  \bibfield  {author} {\bibinfo {author} {\bibfnamefont {Y.-L.}\ \bibnamefont
  {Dong}}, \bibinfo {author} {\bibfnamefont {L.}~\bibnamefont {Chen}}, \bibinfo
  {author} {\bibfnamefont {Y.-J.}\ \bibnamefont {Liu}},\ and\ \bibinfo {author}
  {\bibfnamefont {W.}~\bibnamefont {Li}},\ }\bibfield  {title} {\bibinfo
  {title} {Bilayer linearized tensor renormalization group approach for thermal
  tensor networks},\ }\href {https://doi.org/10.1103/PhysRevB.95.144428}
  {\bibfield  {journal} {\bibinfo  {journal} {Phys. Rev. B}\ }\textbf {\bibinfo
  {volume} {95}},\ \bibinfo {pages} {144428} (\bibinfo {year}
  {2017})}\BibitemShut {NoStop}%
\bibitem [{\citenamefont {Chen}\ \emph {et~al.}(2018)\citenamefont {Chen},
  \citenamefont {Chen}, \citenamefont {Chen}, \citenamefont {Li},\ and\
  \citenamefont {Weichselbaum}}]{Chen2018XTRG}%
  \BibitemOpen
  \bibfield  {author} {\bibinfo {author} {\bibfnamefont {B.-B.}\ \bibnamefont
  {Chen}}, \bibinfo {author} {\bibfnamefont {L.}~\bibnamefont {Chen}}, \bibinfo
  {author} {\bibfnamefont {Z.}~\bibnamefont {Chen}}, \bibinfo {author}
  {\bibfnamefont {W.}~\bibnamefont {Li}},\ and\ \bibinfo {author}
  {\bibfnamefont {A.}~\bibnamefont {Weichselbaum}},\ }\bibfield  {title}
  {\bibinfo {title} {Exponential thermal tensor network approach for quantum
  lattice models},\ }\href {https://doi.org/10.1103/PhysRevX.8.031082}
  {\bibfield  {journal} {\bibinfo  {journal} {Phys. Rev. X}\ }\textbf {\bibinfo
  {volume} {8}},\ \bibinfo {pages} {031082} (\bibinfo {year}
  {2018})}\BibitemShut {NoStop}%
\bibitem [{\citenamefont {{Li}}\ \emph {et~al.}(2023)\citenamefont {{Li}},
  \citenamefont {{Gao}}, \citenamefont {{He}}, \citenamefont {{Qi}},
  \citenamefont {{Chen}},\ and\ \citenamefont {{Li}}}]{tanTRG2023}%
  \BibitemOpen
  \bibfield  {author} {\bibinfo {author} {\bibfnamefont {Q.}~\bibnamefont
  {{Li}}}, \bibinfo {author} {\bibfnamefont {Y.}~\bibnamefont {{Gao}}},
  \bibinfo {author} {\bibfnamefont {Y.-Y.}\ \bibnamefont {{He}}}, \bibinfo
  {author} {\bibfnamefont {Y.}~\bibnamefont {{Qi}}}, \bibinfo {author}
  {\bibfnamefont {B.-B.}\ \bibnamefont {{Chen}}},\ and\ \bibinfo {author}
  {\bibfnamefont {W.}~\bibnamefont {{Li}}},\ }\bibfield  {title} {\bibinfo
  {title} {Tangent space approach for thermal tensor network simulations of the
  2{D} {H}ubbard model},\ }\href
  {https://doi.org/10.1103/PhysRevLett.130.226502} {\bibfield  {journal}
  {\bibinfo  {journal} {Phys. Rev. Lett.}\ }\textbf {\bibinfo {volume} {130}},\
  \bibinfo {eid} {226502} (\bibinfo {year} {2023})}\BibitemShut {NoStop}%
\bibitem [{\citenamefont {Xie}\ \emph {et~al.}(2021)\citenamefont {Xie},
  \citenamefont {Tian}, \citenamefont {Chen}, \citenamefont {Sun},
  \citenamefont {Gao}, \citenamefont {Li}, \citenamefont {Mo},\ and\
  \citenamefont {Shen}}]{Xie2021Giant}%
  \BibitemOpen
  \bibfield  {author} {\bibinfo {author} {\bibfnamefont {H.}~\bibnamefont
  {Xie}}, \bibinfo {author} {\bibfnamefont {L.}~\bibnamefont {Tian}}, \bibinfo
  {author} {\bibfnamefont {Q.}~\bibnamefont {Chen}}, \bibinfo {author}
  {\bibfnamefont {H.}~\bibnamefont {Sun}}, \bibinfo {author} {\bibfnamefont
  {X.}~\bibnamefont {Gao}}, \bibinfo {author} {\bibfnamefont {Z.}~\bibnamefont
  {Li}}, \bibinfo {author} {\bibfnamefont {Z.}~\bibnamefont {Mo}},\ and\
  \bibinfo {author} {\bibfnamefont {J.}~\bibnamefont {Shen}},\ }\bibfield
  {title} {\bibinfo {title} {Giant and reversible low field magnetocaloric
  effect in \ch{LiHoF_4} compound},\ }\href
  {https://doi.org/10.1039/D1DT02958D} {\bibfield  {journal} {\bibinfo
  {journal} {Dalton Trans.}\ }\textbf {\bibinfo {volume} {50}},\ \bibinfo
  {pages} {17697} (\bibinfo {year} {2021})}\BibitemShut {NoStop}%
\bibitem [{\citenamefont {Wendl}\ \emph {et~al.}(2022)\citenamefont {Wendl},
  \citenamefont {Eisenlohr}, \citenamefont {Rucker}, \citenamefont {Duvinage},
  \citenamefont {Kleinhans}, \citenamefont {Vojta},\ and\ \citenamefont
  {Pfleiderer}}]{Wendl2022Mesoscale}%
  \BibitemOpen
  \bibfield  {author} {\bibinfo {author} {\bibfnamefont {A.}~\bibnamefont
  {Wendl}}, \bibinfo {author} {\bibfnamefont {H.}~\bibnamefont {Eisenlohr}},
  \bibinfo {author} {\bibfnamefont {F.}~\bibnamefont {Rucker}}, \bibinfo
  {author} {\bibfnamefont {C.}~\bibnamefont {Duvinage}}, \bibinfo {author}
  {\bibfnamefont {M.}~\bibnamefont {Kleinhans}}, \bibinfo {author}
  {\bibfnamefont {M.}~\bibnamefont {Vojta}},\ and\ \bibinfo {author}
  {\bibfnamefont {C.}~\bibnamefont {Pfleiderer}},\ }\bibfield  {title}
  {\bibinfo {title} {Emergence of mesoscale quantum phase transitions in a
  ferromagnet},\ }\href {https://doi.org/10.1038/s41586-022-04995-5} {\bibfield
   {journal} {\bibinfo  {journal} {Nature}\ }\textbf {\bibinfo {volume}
  {609}},\ \bibinfo {pages} {65} (\bibinfo {year} {2022})}\BibitemShut
  {NoStop}%
\bibitem [{\citenamefont {Liu}\ \emph {et~al.}(2023)\citenamefont {Liu},
  \citenamefont {Yuan}, \citenamefont {Dong}, \citenamefont {Lin},
  \citenamefont {V{\'\i}llora}, \citenamefont {Qi}, \citenamefont {Zhao},
  \citenamefont {Shimamura}, \citenamefont {Ma}, \citenamefont {Wang},
  \citenamefont {Zhang},\ and\ \citenamefont {Li}}]{Liu2023Ultralow}%
  \BibitemOpen
  \bibfield  {author} {\bibinfo {author} {\bibfnamefont {P.}~\bibnamefont
  {Liu}}, \bibinfo {author} {\bibfnamefont {D.}~\bibnamefont {Yuan}}, \bibinfo
  {author} {\bibfnamefont {C.}~\bibnamefont {Dong}}, \bibinfo {author}
  {\bibfnamefont {G.}~\bibnamefont {Lin}}, \bibinfo {author} {\bibfnamefont
  {E.~G.}\ \bibnamefont {V{\'\i}llora}}, \bibinfo {author} {\bibfnamefont
  {J.}~\bibnamefont {Qi}}, \bibinfo {author} {\bibfnamefont {X.}~\bibnamefont
  {Zhao}}, \bibinfo {author} {\bibfnamefont {K.}~\bibnamefont {Shimamura}},
  \bibinfo {author} {\bibfnamefont {J.}~\bibnamefont {Ma}}, \bibinfo {author}
  {\bibfnamefont {J.}~\bibnamefont {Wang}}, \bibinfo {author} {\bibfnamefont
  {Z.}~\bibnamefont {Zhang}},\ and\ \bibinfo {author} {\bibfnamefont
  {B.}~\bibnamefont {Li}},\ }\bibfield  {title} {\bibinfo {title}
  {Ultralow-field magnetocaloric materials for compact magnetic
  refrigeration},\ }\href {https://doi.org/10.1038/s41427-023-00488-7}
  {\bibfield  {journal} {\bibinfo  {journal} {NPG Asia Materials}\ }\textbf
  {\bibinfo {volume} {15}},\ \bibinfo {pages} {41} (\bibinfo {year}
  {2023})}\BibitemShut {NoStop}%
\bibitem [{\citenamefont {Essler}\ and\ \citenamefont
  {Tsvelik}(1998)}]{Essler1998Dynamics}%
  \BibitemOpen
  \bibfield  {author} {\bibinfo {author} {\bibfnamefont {F.~H.~L.}\
  \bibnamefont {Essler}}\ and\ \bibinfo {author} {\bibfnamefont {A.~M.}\
  \bibnamefont {Tsvelik}},\ }\bibfield  {title} {\bibinfo {title} {Dynamical
  magnetic susceptibilities in copper benzoate},\ }\href
  {https://doi.org/10.1103/PhysRevB.57.10592} {\bibfield  {journal} {\bibinfo
  {journal} {Phys. Rev. B}\ }\textbf {\bibinfo {volume} {57}},\ \bibinfo
  {pages} {10592} (\bibinfo {year} {1998})}\BibitemShut {NoStop}%
\bibitem [{\citenamefont {Oshikawa}\ and\ \citenamefont
  {Affleck}(1997)}]{Oshikawa1997Gaps}%
  \BibitemOpen
  \bibfield  {author} {\bibinfo {author} {\bibfnamefont {M.}~\bibnamefont
  {Oshikawa}}\ and\ \bibinfo {author} {\bibfnamefont {I.}~\bibnamefont
  {Affleck}},\ }\bibfield  {title} {\bibinfo {title} {Field-induced gap in
  {$\mathit{S}\phantom{\rule{0ex}{0ex}}=\phantom{\rule{0ex}{0ex}}1/2$}
  antiferromagnetic chains},\ }\href
  {https://doi.org/10.1103/PhysRevLett.79.2883} {\bibfield  {journal} {\bibinfo
   {journal} {Phys. Rev. Lett.}\ }\textbf {\bibinfo {volume} {79}},\ \bibinfo
  {pages} {2883} (\bibinfo {year} {1997})}\BibitemShut {NoStop}%
\bibitem [{\citenamefont {Affleck}\ and\ \citenamefont
  {Oshikawa}(1999)}]{Affleck1999Field}%
  \BibitemOpen
  \bibfield  {author} {\bibinfo {author} {\bibfnamefont {I.}~\bibnamefont
  {Affleck}}\ and\ \bibinfo {author} {\bibfnamefont {M.}~\bibnamefont
  {Oshikawa}},\ }\bibfield  {title} {\bibinfo {title} {Field-induced gap in
  {C}u benzoate and other ${S}=\frac{1}{2}$ antiferromagnetic chains},\ }\href
  {https://doi.org/10.1103/PhysRevB.60.1038} {\bibfield  {journal} {\bibinfo
  {journal} {Phys. Rev. B}\ }\textbf {\bibinfo {volume} {60}},\ \bibinfo
  {pages} {1038} (\bibinfo {year} {1999})}\BibitemShut {NoStop}%
\bibitem [{\citenamefont {Zhao}\ \emph {et~al.}(2003)\citenamefont {Zhao},
  \citenamefont {Wang}, \citenamefont {Xiang}, \citenamefont {Su},\ and\
  \citenamefont {Yu}}]{Zhao2003DMEffect}%
  \BibitemOpen
  \bibfield  {author} {\bibinfo {author} {\bibfnamefont {J.~Z.}\ \bibnamefont
  {Zhao}}, \bibinfo {author} {\bibfnamefont {X.~Q.}\ \bibnamefont {Wang}},
  \bibinfo {author} {\bibfnamefont {T.}~\bibnamefont {Xiang}}, \bibinfo
  {author} {\bibfnamefont {Z.~B.}\ \bibnamefont {Su}},\ and\ \bibinfo {author}
  {\bibfnamefont {L.}~\bibnamefont {Yu}},\ }\bibfield  {title} {\bibinfo
  {title} {Effects of the {D}zyaloshinskii-{M}oriya interaction on low-energy
  magnetic excitations in copper benzoate},\ }\href
  {https://doi.org/10.1103/PhysRevLett.90.207204} {\bibfield  {journal}
  {\bibinfo  {journal} {Phys. Rev. Lett.}\ }\textbf {\bibinfo {volume} {90}},\
  \bibinfo {pages} {207204} (\bibinfo {year} {2003})}\BibitemShut {NoStop}%
\bibitem [{\citenamefont {Lou}\ \emph {et~al.}(2002)\citenamefont {Lou},
  \citenamefont {Qin}, \citenamefont {Chen}, \citenamefont {Su},\ and\
  \citenamefont {Yu}}]{Lou2002Gap}%
  \BibitemOpen
  \bibfield  {author} {\bibinfo {author} {\bibfnamefont {J.}~\bibnamefont
  {Lou}}, \bibinfo {author} {\bibfnamefont {S.}~\bibnamefont {Qin}}, \bibinfo
  {author} {\bibfnamefont {C.}~\bibnamefont {Chen}}, \bibinfo {author}
  {\bibfnamefont {Z.}~\bibnamefont {Su}},\ and\ \bibinfo {author}
  {\bibfnamefont {L.}~\bibnamefont {Yu}},\ }\bibfield  {title} {\bibinfo
  {title} {Field-induced gap in the spin-$\frac{1}{2}$ antiferromagnetic
  {H}eisenberg chain: {A} density-matrix renormalization-group study},\ }\href
  {https://doi.org/10.1103/PhysRevB.65.064420} {\bibfield  {journal} {\bibinfo
  {journal} {Phys. Rev. B}\ }\textbf {\bibinfo {volume} {65}},\ \bibinfo
  {pages} {064420} (\bibinfo {year} {2002})}\BibitemShut {NoStop}%
\bibitem [{\citenamefont {Lou}\ \emph {et~al.}(2005)\citenamefont {Lou},
  \citenamefont {Chen}, \citenamefont {Zhao}, \citenamefont {Wang},
  \citenamefont {Xiang}, \citenamefont {Su},\ and\ \citenamefont
  {Yu}}]{Lou2005Midgap}%
  \BibitemOpen
  \bibfield  {author} {\bibinfo {author} {\bibfnamefont {J.}~\bibnamefont
  {Lou}}, \bibinfo {author} {\bibfnamefont {C.}~\bibnamefont {Chen}}, \bibinfo
  {author} {\bibfnamefont {J.}~\bibnamefont {Zhao}}, \bibinfo {author}
  {\bibfnamefont {X.}~\bibnamefont {Wang}}, \bibinfo {author} {\bibfnamefont
  {T.}~\bibnamefont {Xiang}}, \bibinfo {author} {\bibfnamefont
  {Z.}~\bibnamefont {Su}},\ and\ \bibinfo {author} {\bibfnamefont
  {L.}~\bibnamefont {Yu}},\ }\bibfield  {title} {\bibinfo {title} {Midgap
  states in antiferromagnetic {H}eisenberg chains with a staggered field},\
  }\href {https://doi.org/10.1103/PhysRevLett.94.217207} {\bibfield  {journal}
  {\bibinfo  {journal} {Phys. Rev. Lett.}\ }\textbf {\bibinfo {volume} {94}},\
  \bibinfo {pages} {217207} (\bibinfo {year} {2005})}\BibitemShut {NoStop}%
\bibitem [{\citenamefont {Zvyagin}\ \emph {et~al.}(2004)\citenamefont
  {Zvyagin}, \citenamefont {Kolezhuk}, \citenamefont {Krzystek},\ and\
  \citenamefont {Feyerherm}}]{Zvyagin2004}%
  \BibitemOpen
  \bibfield  {author} {\bibinfo {author} {\bibfnamefont {S.~A.}\ \bibnamefont
  {Zvyagin}}, \bibinfo {author} {\bibfnamefont {A.~K.}\ \bibnamefont
  {Kolezhuk}}, \bibinfo {author} {\bibfnamefont {J.}~\bibnamefont {Krzystek}},\
  and\ \bibinfo {author} {\bibfnamefont {R.}~\bibnamefont {Feyerherm}},\
  }\bibfield  {title} {\bibinfo {title} {Excitation hierarchy of the quantum
  sine-{G}ordon spin chain in a strong magnetic field},\ }\href
  {https://doi.org/10.1103/PhysRevLett.93.027201} {\bibfield  {journal}
  {\bibinfo  {journal} {Phys. Rev. Lett.}\ }\textbf {\bibinfo {volume} {93}},\
  \bibinfo {pages} {027201} (\bibinfo {year} {2004})}\BibitemShut {NoStop}%
\bibitem [{\citenamefont {Oshikawa}\ \emph {et~al.}(1999)\citenamefont
  {Oshikawa}, \citenamefont {Ueda}, \citenamefont {Aoki}, \citenamefont
  {Ochiai},\ and\ \citenamefont {Kohgi}}]{Oshikawa1999}%
  \BibitemOpen
  \bibfield  {author} {\bibinfo {author} {\bibfnamefont {M.}~\bibnamefont
  {Oshikawa}}, \bibinfo {author} {\bibfnamefont {K.}~\bibnamefont {Ueda}},
  \bibinfo {author} {\bibfnamefont {H.}~\bibnamefont {Aoki}}, \bibinfo {author}
  {\bibfnamefont {A.}~\bibnamefont {Ochiai}},\ and\ \bibinfo {author}
  {\bibfnamefont {M.}~\bibnamefont {Kohgi}},\ }\bibfield  {title} {\bibinfo
  {title} {Field-induced gap formation in \ch{Yb_4As_3}},\ }\href
  {https://doi.org/10.1143/JPSJ.68.3181} {\bibfield  {journal} {\bibinfo
  {journal} {Journal of the Physical Society of Japan}\ }\textbf {\bibinfo
  {volume} {68}},\ \bibinfo {pages} {3181} (\bibinfo {year}
  {1999})}\BibitemShut {NoStop}%
\bibitem [{\citenamefont {Kohgi}\ \emph {et~al.}(2001)\citenamefont {Kohgi},
  \citenamefont {Iwasa}, \citenamefont {Mignot}, \citenamefont {F\aa{}k},
  \citenamefont {Gegenwart}, \citenamefont {Lang}, \citenamefont {Ochiai},
  \citenamefont {Aoki},\ and\ \citenamefont {Suzuki}}]{Kohgi2001}%
  \BibitemOpen
  \bibfield  {author} {\bibinfo {author} {\bibfnamefont {M.}~\bibnamefont
  {Kohgi}}, \bibinfo {author} {\bibfnamefont {K.}~\bibnamefont {Iwasa}},
  \bibinfo {author} {\bibfnamefont {J.-M.}\ \bibnamefont {Mignot}}, \bibinfo
  {author} {\bibfnamefont {B.}~\bibnamefont {F\aa{}k}}, \bibinfo {author}
  {\bibfnamefont {P.}~\bibnamefont {Gegenwart}}, \bibinfo {author}
  {\bibfnamefont {M.}~\bibnamefont {Lang}}, \bibinfo {author} {\bibfnamefont
  {A.}~\bibnamefont {Ochiai}}, \bibinfo {author} {\bibfnamefont
  {H.}~\bibnamefont {Aoki}},\ and\ \bibinfo {author} {\bibfnamefont
  {T.}~\bibnamefont {Suzuki}},\ }\bibfield  {title} {\bibinfo {title}
  {Staggered field effect on the one-dimensional ${S}=\frac{1}{2}$
  antiferromagnet \ch{Yb_4As_3}},\ }\href
  {https://doi.org/10.1103/PhysRevLett.86.2439} {\bibfield  {journal} {\bibinfo
   {journal} {Phys. Rev. Lett.}\ }\textbf {\bibinfo {volume} {86}},\ \bibinfo
  {pages} {2439} (\bibinfo {year} {2001})}\BibitemShut {NoStop}%
\bibitem [{\citenamefont {Umegaki}\ \emph {et~al.}(2015)\citenamefont
  {Umegaki}, \citenamefont {Tanaka}, \citenamefont {Kurita}, \citenamefont
  {Ono}, \citenamefont {Laver}, \citenamefont {Niedermayer}, \citenamefont
  {R\"uegg}, \citenamefont {Ohira-Kawamura}, \citenamefont {Nakajima},\ and\
  \citenamefont {Kakurai}}]{Umegaki2015}%
  \BibitemOpen
  \bibfield  {author} {\bibinfo {author} {\bibfnamefont {I.}~\bibnamefont
  {Umegaki}}, \bibinfo {author} {\bibfnamefont {H.}~\bibnamefont {Tanaka}},
  \bibinfo {author} {\bibfnamefont {N.}~\bibnamefont {Kurita}}, \bibinfo
  {author} {\bibfnamefont {T.}~\bibnamefont {Ono}}, \bibinfo {author}
  {\bibfnamefont {M.}~\bibnamefont {Laver}}, \bibinfo {author} {\bibfnamefont
  {C.}~\bibnamefont {Niedermayer}}, \bibinfo {author} {\bibfnamefont
  {C.}~\bibnamefont {R\"uegg}}, \bibinfo {author} {\bibfnamefont
  {S.}~\bibnamefont {Ohira-Kawamura}}, \bibinfo {author} {\bibfnamefont
  {K.}~\bibnamefont {Nakajima}},\ and\ \bibinfo {author} {\bibfnamefont
  {K.}~\bibnamefont {Kakurai}},\ }\bibfield  {title} {\bibinfo {title} {Spinon,
  soliton, and breather in the spin-$\frac{1}{2}$ antiferromagnetic chain
  compound \ch{KCuGaF_6}},\ }\href {https://doi.org/10.1103/PhysRevB.92.174412}
  {\bibfield  {journal} {\bibinfo  {journal} {Phys. Rev. B}\ }\textbf {\bibinfo
  {volume} {92}},\ \bibinfo {pages} {174412} (\bibinfo {year}
  {2015})}\BibitemShut {NoStop}%
\bibitem [{\citenamefont {Shen}\ \emph {et~al.}(2020)\citenamefont {Shen},
  \citenamefont {Zhang}, \citenamefont {Komijani}, \citenamefont {Nicklas},
  \citenamefont {Borth}, \citenamefont {Wang}, \citenamefont {Chen},
  \citenamefont {Nie}, \citenamefont {Li}, \citenamefont {Lu}, \citenamefont
  {Lee}, \citenamefont {Smidman}, \citenamefont {Steglich}, \citenamefont
  {Coleman},\ and\ \citenamefont {Yuan}}]{Bin2020}%
  \BibitemOpen
  \bibfield  {author} {\bibinfo {author} {\bibfnamefont {B.}~\bibnamefont
  {Shen}}, \bibinfo {author} {\bibfnamefont {Y.}~\bibnamefont {Zhang}},
  \bibinfo {author} {\bibfnamefont {Y.}~\bibnamefont {Komijani}}, \bibinfo
  {author} {\bibfnamefont {M.}~\bibnamefont {Nicklas}}, \bibinfo {author}
  {\bibfnamefont {R.}~\bibnamefont {Borth}}, \bibinfo {author} {\bibfnamefont
  {A.}~\bibnamefont {Wang}}, \bibinfo {author} {\bibfnamefont {Y.}~\bibnamefont
  {Chen}}, \bibinfo {author} {\bibfnamefont {Z.}~\bibnamefont {Nie}}, \bibinfo
  {author} {\bibfnamefont {R.}~\bibnamefont {Li}}, \bibinfo {author}
  {\bibfnamefont {X.}~\bibnamefont {Lu}}, \bibinfo {author} {\bibfnamefont
  {H.}~\bibnamefont {Lee}}, \bibinfo {author} {\bibfnamefont {M.}~\bibnamefont
  {Smidman}}, \bibinfo {author} {\bibfnamefont {F.}~\bibnamefont {Steglich}},
  \bibinfo {author} {\bibfnamefont {P.}~\bibnamefont {Coleman}},\ and\ \bibinfo
  {author} {\bibfnamefont {H.}~\bibnamefont {Yuan}},\ }\bibfield  {title}
  {\bibinfo {title} {Strange-metal behaviour in a pure ferromagnetic {K}ondo
  lattice},\ }\href {https://doi.org/https://doi.org/10.1038/s41586-020-2052-z}
  {\bibfield  {journal} {\bibinfo  {journal} {Nature}\ }\textbf {\bibinfo
  {volume} {579}},\ \bibinfo {pages} {51} (\bibinfo {year} {2020})}\BibitemShut
  {NoStop}%
\bibitem [{\citenamefont {Xi}\ \emph {et~al.}(2024{\natexlab{b}})\citenamefont
  {Xi}, \citenamefont {Gao}, \citenamefont {Li}, \citenamefont {Liang},
  \citenamefont {Yu}, \citenamefont {Wang},\ and\ \citenamefont
  {Li}}]{xi2024itp}%
  \BibitemOpen
  \bibfield  {author} {\bibinfo {author} {\bibfnamefont {N.}~\bibnamefont
  {Xi}}, \bibinfo {author} {\bibfnamefont {Y.}~\bibnamefont {Gao}}, \bibinfo
  {author} {\bibfnamefont {C.}~\bibnamefont {Li}}, \bibinfo {author}
  {\bibfnamefont {S.}~\bibnamefont {Liang}}, \bibinfo {author} {\bibfnamefont
  {R.}~\bibnamefont {Yu}}, \bibinfo {author} {\bibfnamefont {X.}~\bibnamefont
  {Wang}},\ and\ \bibinfo {author} {\bibfnamefont {W.}~\bibnamefont {Li}},\
  }\href@noop {} {\bibinfo {title} {Thermal tensor network approach for
  spin-lattice relaxation in quantum magnets}} (\bibinfo {year}
  {2024}{\natexlab{b}}),\ \Eprint {https://arxiv.org/abs/2403.11895}
  {arXiv:2403.11895} \BibitemShut {NoStop}%
\bibitem [{\citenamefont {Weiss}\ and\ \citenamefont
  {Piccard}(1917)}]{Weiss1917}%
  \BibitemOpen
  \bibfield  {author} {\bibinfo {author} {\bibfnamefont {P.}~\bibnamefont
  {Weiss}}\ and\ \bibinfo {author} {\bibfnamefont {A.}~\bibnamefont
  {Piccard}},\ }\bibfield  {title} {\bibinfo {title} {Le ph{\'e}nom{\`e}ne
  magn{\'e}tocalorique},\ }\href
  {https://hal.archives-ouvertes.fr/jpa-00241982/document} {\bibfield
  {journal} {\bibinfo  {journal} {J. Phys. (Paris)}\ }\textbf {\bibinfo
  {volume} {7}},\ \bibinfo {pages} {103} (\bibinfo {year} {1917})}\BibitemShut
  {NoStop}%
\bibitem [{\citenamefont {{A. Smith}}(2013)}]{Smith2013}%
  \BibitemOpen
  \bibfield  {author} {\bibinfo {author} {\bibnamefont {{A. Smith}}},\
  }\bibfield  {title} {\bibinfo {title} {Who discovered the magnetocaloric
  effect? - {Warburg, Weiss,} and the connection between magnetism and heat},\
  }\href {https://doi.org/10.1140/epjh/e2013-40001-9} {\bibfield  {journal}
  {\bibinfo  {journal} {Eur. Phys. J. H}\ }\textbf {\bibinfo {volume} {38}},\
  \bibinfo {pages} {507} (\bibinfo {year} {2013})}\BibitemShut {NoStop}%
\bibitem [{\citenamefont {Xiang}\ \emph {et~al.}(2024)\citenamefont {Xiang},
  \citenamefont {Zhang}, \citenamefont {Gao}, \citenamefont {Schmidt},
  \citenamefont {Schmalzl}, \citenamefont {Wang}, \citenamefont {Li},
  \citenamefont {Xi}, \citenamefont {Liu}, \citenamefont {Jin}, \citenamefont
  {Li}, \citenamefont {Shen}, \citenamefont {Chen}, \citenamefont {Qi},
  \citenamefont {Wan}, \citenamefont {Jin}, \citenamefont {Li}, \citenamefont
  {Sun},\ and\ \citenamefont {Su}}]{Xiang2024Nature}%
  \BibitemOpen
  \bibfield  {author} {\bibinfo {author} {\bibfnamefont {J.}~\bibnamefont
  {Xiang}}, \bibinfo {author} {\bibfnamefont {C.}~\bibnamefont {Zhang}},
  \bibinfo {author} {\bibfnamefont {Y.}~\bibnamefont {Gao}}, \bibinfo {author}
  {\bibfnamefont {W.}~\bibnamefont {Schmidt}}, \bibinfo {author} {\bibfnamefont
  {K.}~\bibnamefont {Schmalzl}}, \bibinfo {author} {\bibfnamefont {C.-W.}\
  \bibnamefont {Wang}}, \bibinfo {author} {\bibfnamefont {B.}~\bibnamefont
  {Li}}, \bibinfo {author} {\bibfnamefont {N.}~\bibnamefont {Xi}}, \bibinfo
  {author} {\bibfnamefont {X.-Y.}\ \bibnamefont {Liu}}, \bibinfo {author}
  {\bibfnamefont {H.}~\bibnamefont {Jin}}, \bibinfo {author} {\bibfnamefont
  {G.}~\bibnamefont {Li}}, \bibinfo {author} {\bibfnamefont {J.}~\bibnamefont
  {Shen}}, \bibinfo {author} {\bibfnamefont {Z.}~\bibnamefont {Chen}}, \bibinfo
  {author} {\bibfnamefont {Y.}~\bibnamefont {Qi}}, \bibinfo {author}
  {\bibfnamefont {Y.}~\bibnamefont {Wan}}, \bibinfo {author} {\bibfnamefont
  {W.}~\bibnamefont {Jin}}, \bibinfo {author} {\bibfnamefont {W.}~\bibnamefont
  {Li}}, \bibinfo {author} {\bibfnamefont {P.}~\bibnamefont {Sun}},\ and\
  \bibinfo {author} {\bibfnamefont {G.}~\bibnamefont {Su}},\ }\bibfield
  {title} {\bibinfo {title} {Giant magnetocaloric effect in spin supersolid
  candidate {Na$_2$BaCo(PO$_4$)$_2$}},\ }\href
  {https://doi.org/10.1038/s41586-023-06885-w} {\bibfield  {journal} {\bibinfo
  {journal} {Nature}\ }\textbf {\bibinfo {volume} {625}},\ \bibinfo {pages}
  {270} (\bibinfo {year} {2024})}\BibitemShut {NoStop}%
\bibitem [{\citenamefont {Li}\ \emph {et~al.}(2024)\citenamefont {Li},
  \citenamefont {Lv}, \citenamefont {Xi}, \citenamefont {Gao}, \citenamefont
  {Qi}, \citenamefont {Li},\ and\ \citenamefont {Su}}]{Li2024TopoCooling}%
  \BibitemOpen
  \bibfield  {author} {\bibinfo {author} {\bibfnamefont {H.}~\bibnamefont
  {Li}}, \bibinfo {author} {\bibfnamefont {E.}~\bibnamefont {Lv}}, \bibinfo
  {author} {\bibfnamefont {N.}~\bibnamefont {Xi}}, \bibinfo {author}
  {\bibfnamefont {Y.}~\bibnamefont {Gao}}, \bibinfo {author} {\bibfnamefont
  {Y.}~\bibnamefont {Qi}}, \bibinfo {author} {\bibfnamefont {W.}~\bibnamefont
  {Li}},\ and\ \bibinfo {author} {\bibfnamefont {G.}~\bibnamefont {Su}},\
  }\bibfield  {title} {\bibinfo {title} {Magnetocaloric effect of topological
  excitations in {Kitaev} magnets},\ }\href
  {https://doi.org/10.1038/s41467-024-51146-7} {\bibfield  {journal} {\bibinfo
  {journal} {Nature Commun.}\ }\textbf {\bibinfo {volume} {15}},\ \bibinfo
  {pages} {7011} (\bibinfo {year} {2024})}\BibitemShut {NoStop}%
\bibitem [{\citenamefont {Li}\ \emph {et~al.}(2020)\citenamefont {Li},
  \citenamefont {Liao}, \citenamefont {Chen}, \citenamefont {Zeng},
  \citenamefont {Sheng}, \citenamefont {Qi}, \citenamefont {Meng},\ and\
  \citenamefont {Li}}]{Li2020}%
  \BibitemOpen
  \bibfield  {author} {\bibinfo {author} {\bibfnamefont {H.}~\bibnamefont
  {Li}}, \bibinfo {author} {\bibfnamefont {Y.~D.}\ \bibnamefont {Liao}},
  \bibinfo {author} {\bibfnamefont {B.-B.}\ \bibnamefont {Chen}}, \bibinfo
  {author} {\bibfnamefont {X.-T.}\ \bibnamefont {Zeng}}, \bibinfo {author}
  {\bibfnamefont {X.-L.}\ \bibnamefont {Sheng}}, \bibinfo {author}
  {\bibfnamefont {Y.}~\bibnamefont {Qi}}, \bibinfo {author} {\bibfnamefont
  {Z.~Y.}\ \bibnamefont {Meng}},\ and\ \bibinfo {author} {\bibfnamefont
  {W.}~\bibnamefont {Li}},\ }\bibfield  {title} {\bibinfo {title}
  {{Kosterlitz-Thouless} melting of magnetic order in the triangular quantum
  {I}sing material \ch{TmMgGaO_4}},\ }\href
  {https://doi.org/10.1038/s41467-020-14907-8} {\bibfield  {journal} {\bibinfo
  {journal} {Nature Commun.}\ }\textbf {\bibinfo {volume} {11}},\ \bibinfo
  {pages} {1111} (\bibinfo {year} {2020})}\BibitemShut {NoStop}%
\bibitem [{\citenamefont {Hu}\ \emph {et~al.}(2020)\citenamefont {Hu},
  \citenamefont {Ma}, \citenamefont {Liao}, \citenamefont {Li}, \citenamefont
  {Ma}, \citenamefont {Cui}, \citenamefont {Shangguan}, \citenamefont {Huang},
  \citenamefont {Qi}, \citenamefont {Li}, \citenamefont {Meng}, \citenamefont
  {Wen},\ and\ \citenamefont {Yu}}]{Hu2020Exp}%
  \BibitemOpen
  \bibfield  {author} {\bibinfo {author} {\bibfnamefont {Z.}~\bibnamefont
  {Hu}}, \bibinfo {author} {\bibfnamefont {Z.}~\bibnamefont {Ma}}, \bibinfo
  {author} {\bibfnamefont {Y.-D.}\ \bibnamefont {Liao}}, \bibinfo {author}
  {\bibfnamefont {H.}~\bibnamefont {Li}}, \bibinfo {author} {\bibfnamefont
  {C.}~\bibnamefont {Ma}}, \bibinfo {author} {\bibfnamefont {Y.}~\bibnamefont
  {Cui}}, \bibinfo {author} {\bibfnamefont {Y.}~\bibnamefont {Shangguan}},
  \bibinfo {author} {\bibfnamefont {Z.}~\bibnamefont {Huang}}, \bibinfo
  {author} {\bibfnamefont {Y.}~\bibnamefont {Qi}}, \bibinfo {author}
  {\bibfnamefont {W.}~\bibnamefont {Li}}, \bibinfo {author} {\bibfnamefont
  {Z.~Y.}\ \bibnamefont {Meng}}, \bibinfo {author} {\bibfnamefont
  {J.}~\bibnamefont {Wen}},\ and\ \bibinfo {author} {\bibfnamefont
  {W.}~\bibnamefont {Yu}},\ }\bibfield  {title} {\bibinfo {title} {Evidence of
  the {Berezinskii-Kosterlitz-Thouless} phase in a frustrated magnet},\ }\href
  {https://doi.org/10.1038/s41467-020-19380-x} {\bibfield  {journal} {\bibinfo
  {journal} {Nature Commun.}\ }\textbf {\bibinfo {volume} {11}},\ \bibinfo
  {pages} {5631} (\bibinfo {year} {2020})}\BibitemShut {NoStop}%
\bibitem [{\citenamefont {Brando}\ \emph {et~al.}(2016)\citenamefont {Brando},
  \citenamefont {Belitz}, \citenamefont {Grosche},\ and\ \citenamefont
  {Kirkpatrick}}]{Brando2016}%
  \BibitemOpen
  \bibfield  {author} {\bibinfo {author} {\bibfnamefont {M.}~\bibnamefont
  {Brando}}, \bibinfo {author} {\bibfnamefont {D.}~\bibnamefont {Belitz}},
  \bibinfo {author} {\bibfnamefont {F.~M.}\ \bibnamefont {Grosche}},\ and\
  \bibinfo {author} {\bibfnamefont {T.~R.}\ \bibnamefont {Kirkpatrick}},\
  }\bibfield  {title} {\bibinfo {title} {Metallic quantum ferromagnets},\
  }\href {https://doi.org/10.1103/RevModPhys.88.025006} {\bibfield  {journal}
  {\bibinfo  {journal} {Rev. Mod. Phys.}\ }\textbf {\bibinfo {volume} {88}},\
  \bibinfo {pages} {025006} (\bibinfo {year} {2016})}\BibitemShut {NoStop}%
\bibitem [{\citenamefont {Gao}\ \emph {et~al.}(2025)\citenamefont {Gao},
  \citenamefont {Huang}, \citenamefont {Maekawa},\ and\ \citenamefont
  {Li}}]{Gao2025SSE}%
  \BibitemOpen
  \bibfield  {author} {\bibinfo {author} {\bibfnamefont {Y.}~\bibnamefont
  {Gao}}, \bibinfo {author} {\bibfnamefont {Y.}~\bibnamefont {Huang}}, \bibinfo
  {author} {\bibfnamefont {S.}~\bibnamefont {Maekawa}},\ and\ \bibinfo {author}
  {\bibfnamefont {W.}~\bibnamefont {Li}},\ }\href
  {https://arxiv.org/abs/2506.22414} {\bibinfo {title} {Spin {Seebeck} effect
  of triangular-lattice spin supersolid}} (\bibinfo {year} {2025}),\ \Eprint
  {https://arxiv.org/abs/2506.22414} {arXiv:2506.22414 [cond-mat.str-el]}
  \BibitemShut {NoStop}%
\bibitem [{\citenamefont {Qu}\ \emph {et~al.}(2024)\citenamefont {Qu},
  \citenamefont {Li}, \citenamefont {Gong}, \citenamefont {Qi}, \citenamefont
  {Li},\ and\ \citenamefont {Su}}]{Qu2024Cuprate}%
  \BibitemOpen
  \bibfield  {author} {\bibinfo {author} {\bibfnamefont {D.-W.}\ \bibnamefont
  {Qu}}, \bibinfo {author} {\bibfnamefont {Q.}~\bibnamefont {Li}}, \bibinfo
  {author} {\bibfnamefont {S.-S.}\ \bibnamefont {Gong}}, \bibinfo {author}
  {\bibfnamefont {Y.}~\bibnamefont {Qi}}, \bibinfo {author} {\bibfnamefont
  {W.}~\bibnamefont {Li}},\ and\ \bibinfo {author} {\bibfnamefont
  {G.}~\bibnamefont {Su}},\ }\bibfield  {title} {\bibinfo {title} {Phase
  diagram, $d$-wave superconductivity, and pseudogap of the {$t$-$t'$-$J$}
  model at finite temperature},\ }\href
  {https://doi.org/10.1103/PhysRevLett.133.256003} {\bibfield  {journal}
  {\bibinfo  {journal} {Phys. Rev. Lett.}\ }\textbf {\bibinfo {volume} {133}},\
  \bibinfo {pages} {256003} (\bibinfo {year} {2024})}\BibitemShut {NoStop}%
\bibitem [{\citenamefont {Fradkin}(2013)}]{fradkin2013field}%
  \BibitemOpen
  \bibfield  {author} {\bibinfo {author} {\bibfnamefont {E.}~\bibnamefont
  {Fradkin}},\ }\href@noop {} {\emph {\bibinfo {title} {Field theories of
  condensed matter physics}}}\ (\bibinfo  {publisher} {Cambridge University
  Press},\ \bibinfo {year} {2013})\BibitemShut {NoStop}%
\bibitem [{\citenamefont {Henkel}(1999)}]{Henkel1999}%
  \BibitemOpen
  \bibfield  {author} {\bibinfo {author} {\bibfnamefont {M.}~\bibnamefont
  {Henkel}},\ }\href@noop {} {\emph {\bibinfo {title} {Conformal Invariance and
  Critical Phenomena}}}\ (\bibinfo  {publisher} {Springer},\ \bibinfo {year}
  {1999})\BibitemShut {NoStop}%
\bibitem [{\citenamefont {Cyuan-Han}\ \emph {et~al.}(2025)\citenamefont
  {Cyuan-Han}, \citenamefont {Vasiliy}, \citenamefont {Rajeev~S.},
  \citenamefont {Alexandre}, \citenamefont {Petr}, \citenamefont {Aike},
  \citenamefont {Matthew~S.}, \citenamefont {David},\ and\ \citenamefont
  {David}}]{3DIsing2025}%
  \BibitemOpen
  \bibfield  {author} {\bibinfo {author} {\bibfnamefont {C.}~\bibnamefont
  {Cyuan-Han}}, \bibinfo {author} {\bibfnamefont {D.}~\bibnamefont {Vasiliy}},
  \bibinfo {author} {\bibfnamefont {E.}~\bibnamefont {Rajeev~S.}}, \bibinfo
  {author} {\bibfnamefont {H.}~\bibnamefont {Alexandre}}, \bibinfo {author}
  {\bibfnamefont {K.}~\bibnamefont {Petr}}, \bibinfo {author} {\bibfnamefont
  {L.}~\bibnamefont {Aike}}, \bibinfo {author} {\bibfnamefont {M.}~\bibnamefont
  {Matthew~S.}}, \bibinfo {author} {\bibfnamefont {P.}~\bibnamefont {David}},\
  and\ \bibinfo {author} {\bibfnamefont {S.-D.}\ \bibnamefont {David}},\
  }\bibfield  {title} {\bibinfo {title} {Bootstrapping the 3d {I}sing stress
  tensor},\ }\href {https://doi.org/https://doi.org/10.1007/JHEP03(2025)136}
  {\bibfield  {journal} {\bibinfo  {journal} {J. High Energ. Phys.}\ }\textbf
  {\bibinfo {volume} {2025}},\ \bibinfo {pages} {136}}\BibitemShut {NoStop}%
\bibitem [{\citenamefont {Xu}\ \emph {et~al.}(2019)\citenamefont {Xu},
  \citenamefont {Sun}, \citenamefont {Lv},\ and\ \citenamefont
  {Deng}}]{Deng2019}%
  \BibitemOpen
  \bibfield  {author} {\bibinfo {author} {\bibfnamefont {W.}~\bibnamefont
  {Xu}}, \bibinfo {author} {\bibfnamefont {Y.}~\bibnamefont {Sun}}, \bibinfo
  {author} {\bibfnamefont {J.-P.}\ \bibnamefont {Lv}},\ and\ \bibinfo {author}
  {\bibfnamefont {Y.}~\bibnamefont {Deng}},\ }\bibfield  {title} {\bibinfo
  {title} {High-precision {Monte Carlo} study of several models in the
  three-dimensional {U}(1) universality class},\ }\href
  {https://doi.org/10.1103/PhysRevB.100.064525} {\bibfield  {journal} {\bibinfo
   {journal} {Phys. Rev. B}\ }\textbf {\bibinfo {volume} {100}},\ \bibinfo
  {pages} {064525} (\bibinfo {year} {2019})}\BibitemShut {NoStop}%
\bibitem [{\citenamefont {Campostrini}\ \emph {et~al.}(2002)\citenamefont
  {Campostrini}, \citenamefont {Hasenbusch}, \citenamefont {Pelissetto},
  \citenamefont {Rossi},\ and\ \citenamefont {Vicari}}]{3DHeisenberg2002}%
  \BibitemOpen
  \bibfield  {author} {\bibinfo {author} {\bibfnamefont {M.}~\bibnamefont
  {Campostrini}}, \bibinfo {author} {\bibfnamefont {M.}~\bibnamefont
  {Hasenbusch}}, \bibinfo {author} {\bibfnamefont {A.}~\bibnamefont
  {Pelissetto}}, \bibinfo {author} {\bibfnamefont {P.}~\bibnamefont {Rossi}},\
  and\ \bibinfo {author} {\bibfnamefont {E.}~\bibnamefont {Vicari}},\
  }\bibfield  {title} {\bibinfo {title} {Critical exponents and equation of
  state of the three-dimensional {H}eisenberg universality class},\ }\href
  {https://doi.org/10.1103/PhysRevB.65.144520} {\bibfield  {journal} {\bibinfo
  {journal} {Phys. Rev. B}\ }\textbf {\bibinfo {volume} {65}},\ \bibinfo
  {pages} {144520} (\bibinfo {year} {2002})}\BibitemShut {NoStop}%
\end{thebibliography}%

$\,$\\
\textbf{Acknowledgements} \\
E.L., Y.J., and W.L. are indebted to Junsen Wang, Xinyang Li, Yuan Gao, Yang Qi and Gang Su for stimulating discussions. This work was supported by the National Key Projects for Research and Development of China (Grant No.~2024YFA1409200), the National Natural Science Foundation of China (Grant Nos.~12222412, 12447101), and the Strategic Priority Research Program of Chinese Academy of Sciences (Grant No. XDB1270000). We thank the HPC-ITP for the technical support and generous allocation of CPU time. 

$\,$\\
\textbf{Author contributions} \\
E.L., Y. J., and W.L. conceived and initiated the research. E.L. and N.X. carried out tensor network calculations. All authors (E.L., N.X., Y.J., and W.L.) conducted data analysis and theoretical analysis. The manuscript was written by E.L. and W.L. with feedback and contribution from all coauthors. W.L. supervised and coordinated the project.

$\,$\\
\textbf{Competing interests} \\
The authors declare no competing interests. 

$\,$\\
\textbf{Additional information} \\
\textbf{Supplementary Information} is available in the online version of the paper. \\
\noindent

\end{document}